\documentclass[preprint,12pt, dvipsnames, usenames]{elsarticle}

\usepackage{amssymb}
\usepackage{graphicx, subfigure}
\usepackage{amsmath}
\usepackage{tikz}
\usepackage{natbib}
\usepackage{caption}
\usepackage{subcaption}
\usetikzlibrary{arrows,decorations.pathmorphing,backgrounds,positioning,fit,petri}
\usetikzlibrary{shapes.geometric}
\usetikzlibrary{decorations.pathreplacing}
\usepgflibrary{shapes.symbols}
\usetikzlibrary{circuits.ee.IEC}
\usetikzlibrary{circuits.logic}
\usetikzlibrary{circuits.logic.IEC}
\usepackage{pgfplots}
\usetikzlibrary{decorations.text}
\usepackage{amsfonts}
\usepackage{array}
\biboptions{sort&compress}
\usepackage{soul}
\usepackage{amsthm}
\usepackage{placeins}
\usepackage{cprotect}
\usepackage{geometry}
\newcommand{\CPU}[2]
{   \begin{scope}[shift={(#1,#2)}]
   \draw [fill=Blue!70] (-0.5,-0.5) rectangle (0.5,0.5); 
   \draw [fill=Goldenrod!45] (-0.25,-0.25) rectangle (0.25,0.25);
   \node at (0,0.8) {CPU};
  \end{scope}
}
\newcommand{\PC}[2]
{
  \begin{scope}[shift={(#1,#2)}]
  \draw [ultra thick, draw=black, fill=gray!100] (-2,-1) -- (-2,1) -- (2,1) -- (2,-1) -- (0.4,-1) -- (0.4,-1.4) -- (0.8,-1.4)--(1.2,-1.8)--(-1.2,-1.8)--(-0.8,-1.4) -- (-0.4,-1.4) -- (-0.4,-1)-- cycle;
  \draw [fill=Blue!40] (-1.8, -0.9) rectangle (1.8,0.9);
  \node at (0,0) {\Huge \textcolor{White}{NELM}};
  \CPU{-2}{2};
  \CPU{-0.67}{2};
  \CPU{0.67}{2};
  \CPU{2}{2};
  \end{scope}
}
\newcommand{\CPE}[0]
{~~~~~~~~~~~\begin{tikzpicture}
   \node at (0,0) [draw=black, fill=white, thick] {CPE};
   \draw [thick](-0.775,0)--(-0.55,0);
   \draw [thick](0.775,0)--(0.55,0);   
 \end{tikzpicture}
}
\newcommand{\CPEabstract}[0]
{
~~~~~~~~~~~\begin{tikzpicture}
   \node at (0,0) [draw=black, fill=OliveGreen, thick] { \textcolor{white} {CPE}};
   \draw [thick](-0.775,0)--(-0.55,0);
   \draw [thick](0.775,0)--(0.55,0);   
 \end{tikzpicture}
}

\newcolumntype{C}[1]{>{\centering\let\newline\\\arraybackslash\hspace{0pt}}m{#1}}

\journal{Applied Soft Computing}

\begin{document}

\begin{frontmatter}

\title{NELM: Modern Open-Source Software for Multipurpose Impedance Spectra Analysis}

\author[label1]{Natalia A. Boitsova}
\ead{natab2002@yandex.ru}
\author[label1]{ Anna A. Abelit}
\author[label1]{ Daniil D. Stupin}

\affiliation{[label1]organization={Alferov University},
            addressline={Khlopina st., 8/3}, 
            city={St. Petersburg},
            postcode={194021}, 
            country={Russian Federation}}

\begin{abstract}
Nowadays electrical impedance spectroscopy (EIS) has become an advanced experimental technique with a wide range of applications: from simple passive circuits diagnostics to semiconductor high-end device development and breakthrough technologies in bio-sensing. Although hardware for EIS today is well developed,  the EIS analysis software is mainly custom, old fashioned, \textit{i.e.} it is limited by features, does not utilize the progress in the modern computer science and hardware, and is usually implemented in close-source code or written on outdated programming languages, which causes slow progress in  field of the EIS and complicates researchers attempts of development  in EIS  autonomous devices, such as implants.
In this article, we introduce a free and open-sourced \textsc{MatLab/GNU Octave} package for EIS analysis called NELM, which provides powerful equipment tools for matching experimental impedance data with theoretical equivalent circuits. Our software has an user friendly interface and supports different formats of input data, fitting programs, and impedance models. In addition, we have developed NELM with implementation of the latest progress in computation science such as symbolic calculations, parallel computing, and artificial intelligence.  
The abilities of NELM were validated by its applications in the different fields of science, such as semiconductor studies, bioimpedance and electrochemestry, which demonstrated high-efficiency of the proposed software package and showed that it is a promising tool for solving actual problems in electronic industry, biosensorics, and healthcare technologies. 
\end{abstract}

\begin{graphicalabstract}

\begin{center}
\hspace{-1 cm}
\begin{tikzpicture}[scale=1,circuit ee IEC, thick]

\draw[draw=Green!75, thin] (-3.5, -6) -- (3.5, -6);
\draw[draw=Green!75, thin] (-3.5, 5.7) -- (3.5, 5.7);
\draw[draw=Green!75, fill=Green!20] (-6.5, 6.5) rectangle (9.3, -6);
\draw[draw=Green!75, fill=Green!95] (-6.5, 6.5) rectangle (9.3, 5.7);
\draw[draw=Green!20, fill=Green!10] (-3, 5.7) rectangle (3, -6);
\draw[draw=Green!75] (-3.5, -6) -- (3.5, -6);
\draw[draw=Green!75] (-3.5, 5.7) -- (3.5, 5.7);
\draw[draw=Green!75] (-6.5, 6.5) rectangle (9.3, -6);
\draw[draw=Green!75] (-6.5, 6.5) rectangle (9.3, 5.7);
  \begin{scope}[shift={(1.15, 0)}]
       \begin{scope}[shift={(-6, 3)}]
    \node at (0,0) {\includegraphics[scale=0.25]{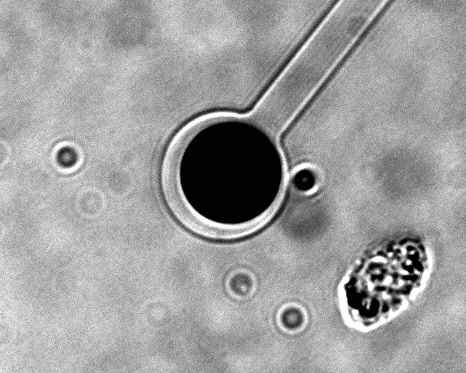}};
    \def\x{-0.3} 
    \def\y{-0.5}
    \draw [-stealth,  Yellow, ultra thick] (1.25+\x,-0.27+\y)--(1+\x,0+\y);
\end{scope}
\begin{scope}[shift={(-6, 1)}]
    \node at (0,0) {\includegraphics[scale=0.25]{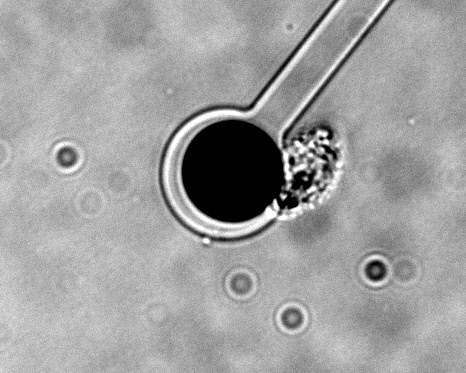}};
    \def\x{-0.6} 
    \def\y{-0.08}
    \draw [-stealth,  Yellow,  ultra thick] (1.3+\x,-0.28+\y)--(1+\x,0+\y);
\end{scope}
\begin{scope}[shift={(-6, -1)}]
    \node at (0,0) {\includegraphics[scale=0.25]{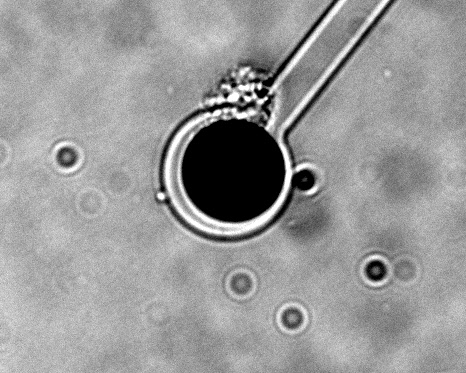}};
     \def\x{-0.8} 
    \def\y{0.3}
    \draw [-stealth,  Yellow,  ultra thick] (1.3+\x,-0.28+\y)--(1+\x,0+\y);
\end{scope}
\begin{scope}[shift={(-6, -3)}]
    \node at (0,0) {\includegraphics[scale=0.25]{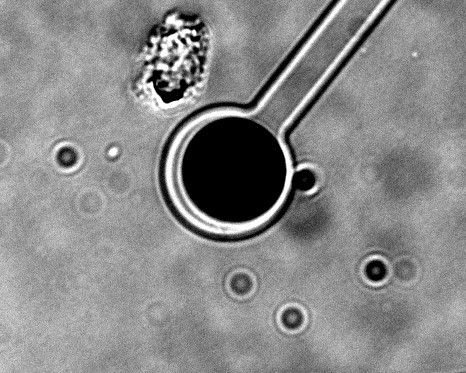}};
   \def\x{-1} 
    \def\y{0.5}
    \draw [-stealth, Yellow,  ultra thick] (1.3+\x,-0.28+\y)--(1+\x,0+\y);
\end{scope}
  \end{scope}

\begin{scope}[shift={(1.15, 0)}]
    \draw [Gray!70, ultra thick] (-5, 3)--(-4.5, 3);
    \draw [Gray!70, ultra thick] (-5, 1)--(-4.5, 1);
    \draw [Gray!70, ultra thick] (-5, -1)--(-4.5, -1);
    \draw [Gray!70, ultra thick] (-5, -3)--(-4.5, -3);
    \draw [Gray!70, ultra thick](-4.5, 3)--(-4.5, -3);
\end{scope}
\draw [->, Gray!100, ultra thick] (-3.35, 0)--(-2.05, 0);
\draw [->, Blue!75, ultra thick] (2, 0)--(3.3, 0);
\draw [<-, Blue!75, ultra thick] (-0, -1.9)--(-0, -4.5) node[midway, above, rotate=90] {\Large Model};
\draw [<-, Blue!75, ultra thick] (0, 3)--(0, 5);
  \draw[draw=Blue!75, fill=Blue!20] (3.3, 3.7) rectangle (8.7, -3.7);
\node at (6,-1.7) {\includegraphics[scale=0.17]{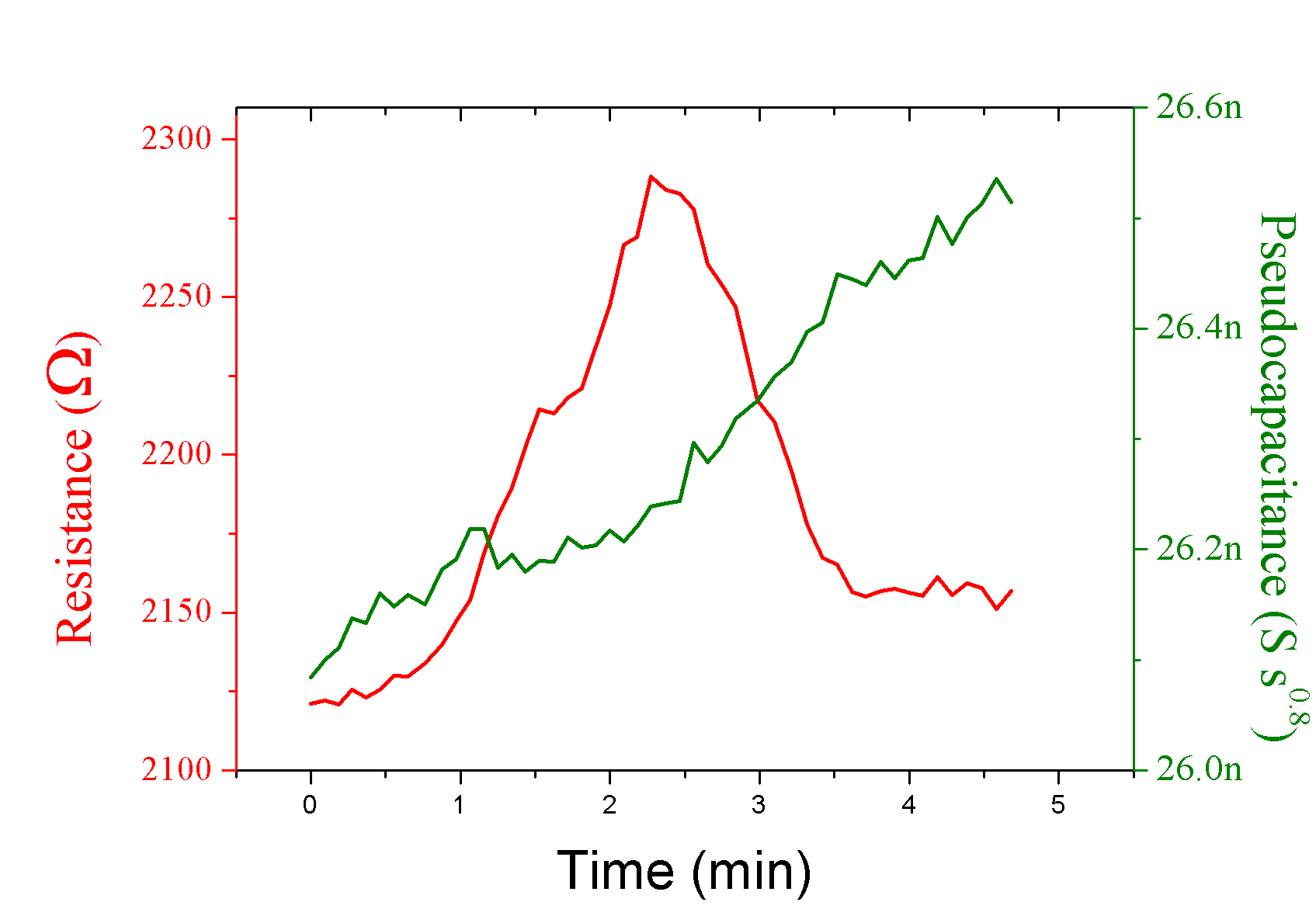}};
\node at (6,1.7) {\includegraphics[scale=0.17]{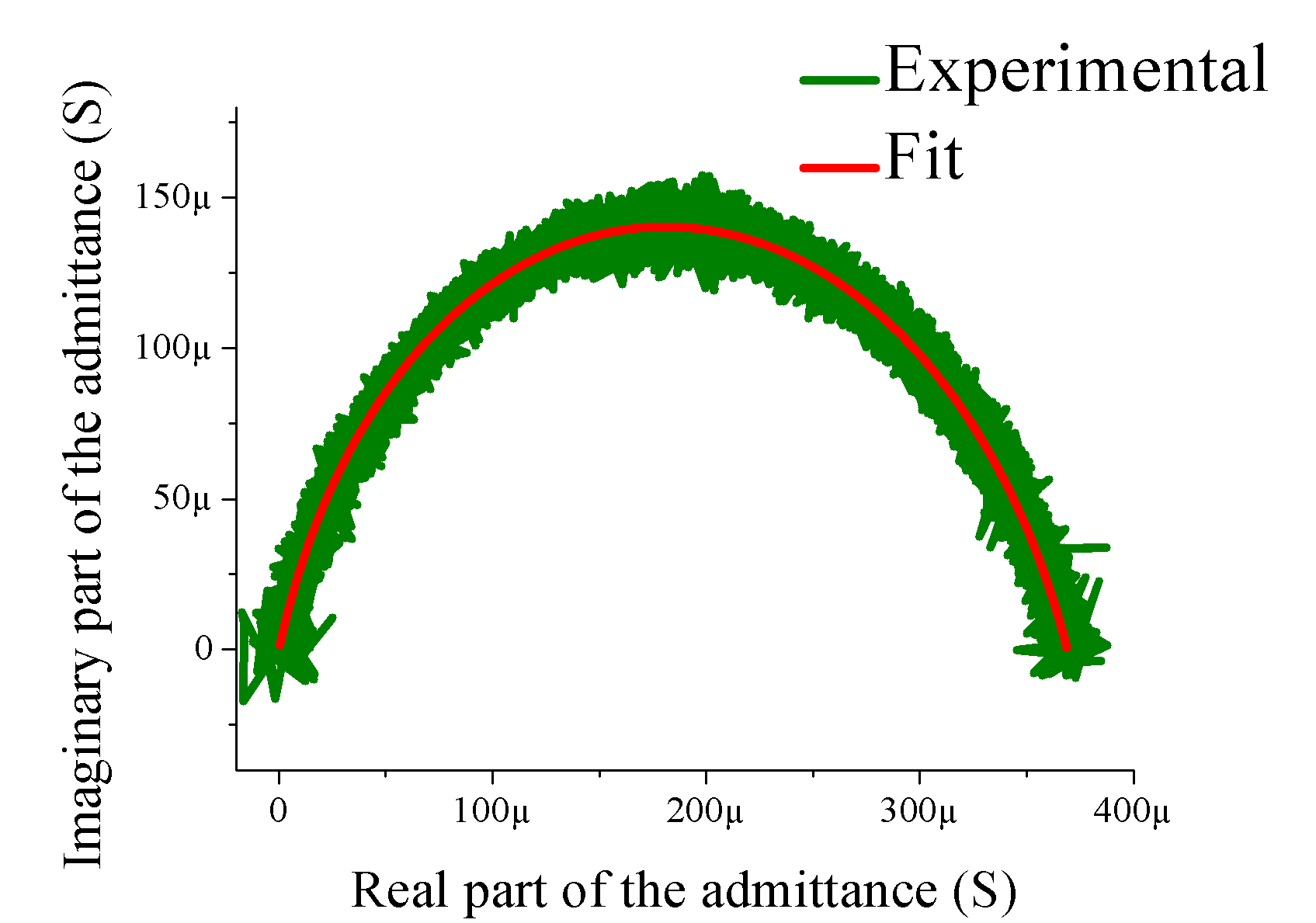}};
\draw[draw=Blue!75, fill=Blue!20] (-1,4)--(1,4)--(1,5.5)--(-0.75,5.5)--(-1,5)--(-1,4);
\draw[draw=Blue!75, fill=Blue!20] (-0.75,5.5)--(-0.75,5)--(-1,5);
\node at (0,4.725)  {\textcolor{RoyalBlue}{Settings}};
\node at (-4.85, 6){\textcolor{White}{\large Experiment}};
\node at (0, 6){\textcolor{White}{\large Data analyze}};
\node at (6, 6){\textcolor{White}{\large Phenomena discover }};
   \PC{(0, 0)};
 \draw[draw=Blue!75, fill=Blue!20, rounded corners=2mm] (-2.8, -4.5) rectangle (2.8, -5.5);

\begin{scope}[shift={(-1,-5)}]
   \draw [Red!50] (1.6,0)  -- (1.45+1.55,0);
    \draw (-1.5,0) to[contact={color=Blue}] (-1.5,0) -- (-1.4,0) to[resistor={color=Red}]  (0.1,0) to[inductor={color=violet}] (1.6,0)  node {\CPEabstract} (1.45+1.55,0) -- (1.6+1.55+0.2,0) to[contact={color=Blue}] (1.6+1.55+0.2,0) ;
     \end{scope}
\end{tikzpicture}

\end{center}
\end{graphicalabstract}

\begin{highlights}
\item Multipurpose open-source software NELM for impedance spectra analyzing was designed
\item NELM was written in well-tuned \textsc{MatLab} and is supported by the power of the parallel and symbolic computing 
\item Program efficiency was demonstrated on the different scientific tasks: from semiconductors to biology
\end{highlights}


\end{frontmatter}

\tableofcontents
\section{Introduction}
\label{sec:Intro}

Typical experimental science practices include two essential steps: data collection and data processing \cite{stupin2021bioimpedance,oppenheim1976digital, Tietze,DSP_Book_Dimitris,DSP_Book_Rawlings,DSP_Book_Yan,Statistics_book,Hansen}. The final aim of the last step usually is to find the relationships in the experimental data, which can be achieved by matching it to some $ab\,initio$ theory or to appropriate phenomenological models. For impedance spectroscopy, both theoretic and phenomenological models can often be represented in terms of equivalent circuits (EC), electrical schemes that have the same impedance as the sample under test \cite{Barsoukov, macdonald1982applicability,MacdonaldR}. The advantage of such an approach consists in the possibility of separating in the observed spectra the phenomena with a distinct nature, because frequently different processes in the investigated samples can be attributed to certain elements in an equivalent circuit of the sample \cite{stupin2021bioimpedance, Barsoukov}. Thus, fitting of the equivalent scheme parameters in order to match its impedance spectrum to the experimentally obtained spectrum provides a promising and powerful tool for the investigation of complex systems, including biological objects \cite{stupin2021bioimpedance,Stupin2017,rahman2009detailed,qiu2008real,qiu2009intervention}, chemical sensors \cite{ganguly2019passively,ismail2024facile,zarei2022impedimetric,shervedani2006novel}, and semiconductor devices \cite{lee2021two,crain2013comparison,shervedani2006novel}.

For implementation of this EIS processing type, there are three common approaches: algebraic, geometric, and complex non-linear least squares (CNLS) \cite{stupin2021bioimpedance, macdonald1963cnls, MacdonaldD, Tsai,MacdonaldR,macdonald1982applicability}. The algebraic approach consists of the evaluation of EC parameters using spectrum features (minima, maxima, peak position values, $etc$). The geometric approach uses geometric manipulations with impedance spectra plots for estimation of EC parameters. The CNLS approach, introduced by J. Ross Macdonald in the 1960s \cite{macdonald1963cnls}, today is the most common and accurate method for EC reconstruction, because it can be applied to a wide range of EC topologies and because it uses the whole measured spectrum points in contrast to algebraic and geometric methods. Due to these reasons, the NELM software is basically developed for producing CNLS approximations of the experimental immittance spectra.

The key idea of the CNLS method, which we illustrate in the case of admittance representation,  is to minimization of the functional 
\begin{equation}
\label{eq:Working_Function}
    E^2=\dfrac{1}{L}\sum\limits_{\ell=1}^{L}|Y_{m}(\omega_\ell)-Y_{ex}(\omega_\ell)|^2\times \mathcal{W}(\omega_\ell)=\min,
\end{equation}
where $Y_{ex}(\omega_\ell)$ is experimentally obtained admittance at the frequency $\omega_\ell$, $Y_m(\omega_\ell)$ is the EC model admittance at the same frequency, $L$ is the total spectrum measured points, and $E^2$ is the mean squared error, or so-called working function in the terms of optimization theory, $\mathcal{W}(\omega_\ell)\geq 0$ is the weight of the data-point at $\omega_\ell$ frequency. Instead, the admittance $Y$ in Eq.~\eqref{eq:Working_Function} can be used with any immittance value, such as impedance, impedance magnitude, $etc$ (see Sec.~\ref{sec:state_of_the_art_proc}). Because typically the EC impedance depends on the ECs' parameters in the indirect, non-linear manner, for solving Eq.~\eqref{eq:Working_Function} was used the numerical optimization methods, the most popular of which are gradient descent, Gauss-Newton, Levenberg-Marquardt, and Nelder-Mead \cite{nocedal1999numerical}. 
The earliest known wide-spread software for providing CNLS impedance analysis is the famous LEVM program by J. Ross Macdonald, which is often used today, despite its limitations and the fact that it is technically outdated \cite{LEVM_manual}. 

One of the commonly used today commercial software for EIS is EC-lab by BioLogic (France) \cite{EC_Lab_manual}, which is usually connected to a BioLogic device, allowing users to analyze data in the same program that they used to gather it. This program version 11.50 has a free demo shareware, and it is possible to transfer users' data from outside devices with a function import from ASCII format. This package also operates with a user-friendly interface that constructs equivalent circuits from different elements, but their selection is limited to parallel/series connections, $e.i.$ bridge schemes not supported by EC-Lab. In addition, the schemes in EC-Lab are constructed using lumped elements, and thus not all immittance phenomena can be taken into account during CNLS processing in this package. To be specific, $e.g.$ it is impossible in EC-Lab add to immittance delay exponent multiplier  $e^{-i \omega \tau}$, which appear, in particular, due to the time offset between current and voltage measurements using multichannel ADC or when a non-ideal cable is used  \cite{naranjo2024smart}. 
In EC-Lab, users can choose a method of approximation from the five provided, regulate the number of iterations, and even use weights, but only as $\mathcal{W}=|Z|^-1$. We also found that EC-Lab has performed perfect approximation of model data of RLC circuit, but did not manage to succeed in the case of experimental data, possibly due to the noise and inability to add the mentioned above delay exponent to the model. 

Another example of frequently used software, which deals with CNLS, is commercially available ZView by Scribner Associates Inc. ZView version 4.0 is partly based on LEVM and has a relatively modern, user-friendly graphical interface, and also provides a wide range of EC choosing \cite{ZView_manual}. It also includes an instant fit feature, which is likely based on the EC algebraic search method. However, since ZView is provided as a Windows executable file with closed source code,  it has lower task flexibility  compared to LEVM,  source code of which can be modified for ongoing impedance research. In addition, unlike LEVM, ZView 4.0 supports batch processing of spectra, which, however, is implemented just in a pipelined manner, $e.g.$ no refinements are used to initialize the CNLS for time series impedance data (cmp. with Sec.~\ref{sec:MC} and \cite{Boitsova2024MC}).
Unfortunately, ZView has a limitation in visualizing impedance spectra at high resolution (spectrum plots only display 108 data points, possibly due to optimization of video outputs), which makes preliminary analysis of spectra difficult. In contrast, LEVM does not support large, high-resolution spectra. Finally,  EC-Lab, LEVM, and ZView, by default, are not compatible with parallel processing and modern high-speed impedance spectroscopy approaches such as adaptive filtering EIS and Fourier-EIS \cite{stupin2021bioimpedance}.  

In recent years, several custom \cite{cavallini2019ecis} and open-source programs for processing impedance spectra \cite{Yrjana2022} have also appeared. In particular, the \textsc{Python} packages are very popular today \cite{Yrjana2022,Impedance_py,Impedance_py_2,pimpspec,Py_EIS,ImpedanceAnalyzer,EIS_Spectrum_Analyser_Z_only,AutoEIS_article,AutoEIS_source,OSIF,pyZwx_article,pyZwx_manual,maghsoudi2018matlab}, however, they, like other open-source products in one combination or another have the same limitations as EC-Lab, Zview and LEVM, $e.g.$ primary parallel/series EC type concept (usage the Circuit Description Code \cite{boukamp1986nonlinear}), usage of only built-in optimization methods, like \verb!lmfit! \textsc{Python} package, only impedance ($Z$) fitting is released, no hardware acceleration by default, lack of  time-domain EIS support, labor-intensive installation and use, $etc.$ 

Trying to eliminate the limitations of existing state-of-the-art programs for impedance phenomena analysis and inspired by the LEVM program, in this study, we have developed the open-source \textsc{MatLab/GNU Octave-based} software called NELM, which includes the advantages of the LEVM, EC-Lab, and ZView, and, in addition, provides rich and flexible opportunities for challenging CNLS-problems mentioned above. The alpha version of NELM (NELder-Mead acronym) was initially introduced in 2017 \cite{Stupin2017} as a request-available software, which our team has used for bioimpedance data analysis \cite{Stupin2017,stupin2018take,stupin2021bioimpedance}. 
In this study, we have significantly improved and optimized its source code, which can be download via link \cite{NELM_download} at GitHub.
We have aimed to make an open-source software package with a clear user-friendly structure that allows people with minimal programming experience  to be able to produce CNLS analysis with different equivalent schemes, minimization protocols, and miscellaneous data processing precisely for their needs. The purpose of the current article, which can be viewed as a manual, is to introduce our program to new potential users, guide them through their use of NELM and give some possible application of this package including analysis of the passive circuits, semiconductor devices, electrochemistry, and biosensing.

This paper is organized as follows. First, there is brief information about the CNLS approximation and the theory behind the minimization strategies. After that, there is an overview of the structure of the program and a short manual for users. Following that we introduce results of the different NELM performed approximations for various experiments, such as the RLC circuit identification, HeLa cell research using a multielectrode array, photodiode characterization, and EIS spectra noise canceling example. Finally, some outcomes and further development of NELM are presented and discussed.

\section{NELM structure, ``phrase-book'', and ``guidebook''}
Since most of the article shows interconnections between the NELM source code and EIS theory and experimental techniques, here we introduce a small vocabulary, provided in order to explain the meaning of NELM essential variables and routines. We have used in the text the \verb!verbatim! font to indicate the programming variables of the source code file names. 

In total, NELM consists of five parts: three interfaces for data manipulation and NELM tweaking, the core for data analysis, and the interface for result storage (Fig.~\ref{fig:NELM_structure}). The key NELM variables are listed on the Tab.~\ref{tab:Essential_var}. Several NELM adjustments, which were created to give user an ability to customize (sound notification or batch processing mode), can be triggered by changing values of the variables, which are highlighted at the beginning of the main \verb!NELM.m! file by 
\begin{center}
\verb!%---->>>>!    
\end{center} comment. These tweaks are devoted for user comfort, and are not critical for main data analysis. Contrary, the tweaks which are essential for NELM signal processing (model type, minimization algorithm, $etc.$) are stored in special m-files, which has file-name mask \verb!Settings_*.m!. The important tweaks in these files are denoted by fish-like comment\begin{center}
\verb!%>---]}! 
\end{center}   The loading of the settings file is provided via a special graphical interface. We have used such separation of the tweaks to provide NELM flexibility, since the first of them allows to convert immittance data from arbitrary format to the format understandable for NELM, while the second part of tweaks are used to control the NELM CNLS data-processing. The settings essential for CNLS processing such as minimization methods, frequency and time range, $etc.$ are stored in the \verb!options! structure. Since NELM was initially developed for analyzing high-resolution multichannel EIS  data,  we introduce  the structure \verb!Channels!, which contains the information about NELM processing results in the standardized form.  To provide the flexibility for the fitting model choice NELM program uses the anonymous function \verb!Channels.Model! as an interface. It allows the user to specify the arbitrary model types in the format understandable by NELM. 
Thus, NELM operator can use the different processing techniques for different data types. 

In summary, the NELM operates as follows: the input data is converted into the NELM-compatible format by \verb!@Get_Spectrum_Func! interface, then the data is processed by \verb!NELM_Func! core in accordance with \verb!Channels.Model! and \verb!options! interfaces, and then the NELM output is stored into \verb!Channels! structure.
\begin{table}[]
    \centering
    \footnotesize
    \begin{tabular}{p{5 cm} p{8 cm}}
    \hline
    Variable or routine& Description \\ \hline \hline
    \verb!NELM.m!    & The main script, by running which user starts process of approximation.    \\ \hline
    \verb!NELM_Func.m!  & The main routine, which carries out the CNLS approximation.\\  \hline 
    \verb!options! & The structure which controls NELM core execution. \\ \hline
    \verb!N! & Total number of the Monte-Carlo iterations. \\ \hline
    \verb!Get_Spectrum_Func! & The data loading interface anonymous function.\\ \hline
    \verb!Channels! & The structure for storage NELM output.\\ \hline 
    \hline
    \end{tabular}
    \caption{Description of the essential NELM variables and routines.}
    \label{tab:Essential_var}
\end{table}
 \begin{figure*}
 \label{fig:guidebook}
    \centering
\begin{tikzpicture}
    \node at (0,4) [fill=green!25, draw=green!75, rounded corners=2 mm]{Data};
    \node at (0,2) [text width =4 cm, fill=Plum!25, rounded corners=2mm] {\footnotesize @Get\_Spectrum\_Func\\ Data loading interface };
    \node at (0,0) [text width = 4 cm, fill=Red!50, rounded corners=2mm]{\centering \footnotesize NELM\_Func\\ Approximation routine};
    \node at (0,-2) [text width = 4 cm, fill=Sepia!50, rounded corners=2mm]{\centering \footnotesize Channels\\ Structure with results};
    \node at (-4,0) [text width = 5 cm, rotate=90, fill=Yellow!25, rounded corners=2mm]{\centering \footnotesize Channels.Model\\ Equivalents scheme interface};     
    \node at (4,0) [text width = 5 cm, rotate=90,fill=SkyBlue!25, rounded corners=2mm]{\centering \footnotesize options\\ \scriptsize Structure with approximation tweaks};   
     \draw[thick,->] (0,3.7) -- (0,2.55);
     \draw[thick,->] (0,1.45) -- (0,0.55);
     \draw[thick,->] (0,-0.55) -- (0,-1.45);
     \draw[thick,->] (-3.5, 0) -- (-2.2, 0);
     \draw[thick,->] (3.5, 0) -- (2.2, 0);
\end{tikzpicture}
    \cprotect\caption{Architecture of the \textsc{NELM} engine. To provide the program flexibility NELM has: data loading interface, which convert the data in arbitrary format  to the \textsc{MatLab} complex column-array by anonymous function \verb!@Get_Spectrum_Func!; model loading interface, which evaluate by  anonymous function \verb!Channels.Model! the EC immittance; the structure \verb!options!, which define the approximations process settings; the main universal approximation core \verb!NELM_Func!; the mentioned above structure \verb!Channels!, which store approximation results, including EC parameters, model options, model type, data file names, $etc$.}
    \label{fig:NELM_structure}
\end{figure*}
\begin{figure*}
\vspace{-1cm}
\hspace{-1cm}
\begin{tikzpicture}
 \begin{scope}[scale=0.9, shift={(5,10)}]
  \node at (-0.3,0) [align=center,  draw=Gray!50] {\scriptsize \verb!NELM.m!};
  
  \begin{scope}[shift={(-2.1,0.25)}]
      \draw [very thick, draw=Green, rounded corners=1.5mm,->] (-0.3,-0.2)--(0.1, -0.2);      
  \end{scope}
 
  \node at (3.8,0.75) {\tiny Manual initializing of the:};
  \node at (5,0) [draw=Goldenrod!50, fill=Goldenrod!20]{\tiny \begin{tabular}{p{1.8 cm}|p{2.8cm}}
                          Batch mode   & \verb!Manual_Batch!\\
                          Files        & \verb!file_names!\\
                          Data format  & \verb!I_select_file_type_number!\\
                        \end{tabular}};
 \end{scope}
  \begin{scope}[scale=0.9, shift={(4.8,7.5)}]
  \node at (5.5,1.6)  {\tiny Automatic loading from selected \verb!Settings_*.m! file:}; 
  \node at (-0.1,1) [draw=Gray!50] {\tiny \verb!Selecting_Settings_GUI.m!};
  \begin{scope}[shift={(-1.9,1.25)}]
      \draw [very thick, draw=Green, rounded corners=1.5mm,->] (-0.2,0.2) -- (-0.2,-0.2)--(0.1, -0.2);      
  \end{scope}

  \node at (-0.1,-0.25) [draw=Gray!50] {\scriptsize \verb!NELM.m!};
    \begin{scope}[shift={(-1.9,-0.4)},yscale=-1,xscale=1]
      \draw [very thick, draw=Green, rounded corners=1.5mm,->] (-0.2,0.2) -- (-0.2,-0.2)--(0.1, -0.2);      
  \end{scope}
  \node at (5.5,-0.2) [draw=Goldenrod!50, fill=Goldenrod!20]{\tiny \begin{tabular}{p{2.1 cm}|p{3.2 cm}}
                          CNLS solver               & \verb!options.method!\\
                          CNLS Model                & \verb!Channels.Model!\\
                          Fixed parameters          & \verb!Channels.Model_Options.fix_pars!\\
                          Starting EC ratings       & \verb!Channels.x0_init!\\
                          MC iterations number      & \verb!N!\\
                          Warming up mode           & \verb!Is_Warming_Up!\\
                          MC mode                   & \verb!options.Monte_Carlo_Mode!\\
                          Data files range          & \verb!Start_From! \& \verb!Finish_At!\\               
                        \end{tabular}};
 \end{scope}

 \begin{scope}[scale=0.9, shift={(0,0)}]
  \draw [rounded corners=2mm, ultra thick, ->,draw=ProcessBlue] 
        (-1,10)--(-1,8.9);
  \draw [rounded corners=2mm, ultra thick, ->,draw=ProcessBlue] 
        (-1,8.5)--(-1,7.8);  
  \draw [rounded corners=2mm, ultra thick, ->,draw=ProcessBlue] 
        (-1,6.8)--(-1,6.05);   
  \draw [rounded corners=2mm, ultra thick, ->,draw=ProcessBlue] 
        (2.2,-7.9)--(2.2,-8.5);  
  \draw [rounded corners=2mm, ultra thick, ->,draw=ProcessBlue] 
        (2.2,-9.5)--(5.75,-9.5);  
  \draw [rounded corners=2mm, ultra thick, ->,draw=ProcessBlue] 
        (2.2,-9.5)--(-5.5,-9.5) -- (-5.5,5.3) -- (-3.2,5.3); 
  \node at (5,-9.2)  {Yes}      ;
  \node at (-0.8,-9.2)  {No}      ;  
    \node at (-1.0,10) [ text width=5.0 cm, draw=ProcessBlue!75, fill=ProcessBlue!20, rounded corners=2mm, font=\tiny] {\baselineskip=1mm \scriptsize Starting \textsc{MatLab}, premiliar NELM adjustment, and connecting to data storage};
    \node at(-1.0,8.525) [align=center, text width=5.0 cm, draw=Cerulean!75, fill=Cerulean!20, rounded corners=2mm] {\scriptsize Loading settings for CNLS analysis};
    \node at(-1.0,7.1) [align=center, text width=5.0 cm, draw=Green!75, fill=Green!20,trapezium,
                        trapezium left angle = 65,
                        trapezium right angle = 115,
                        trapezium stretches, font=\tiny] {\scriptsize Loading experimental data $Y_{ex}$ via \verb!@Get_Spectrum_Func! and initializing approximation parameters};
    \begin{scope}[shift={(-1.0,1.5)}]
       \draw [rounded corners=2mm, ultra thick, draw=ProcessBlue,->] 
             (2,3.8) -- (3.5,3.8) node [above, midway] {\tiny Yes}; 
       \draw [rounded corners=2mm, ultra thick, draw=ProcessBlue,->] 
            (0,3.2) -- (0,2.6);
       \node at (-0.4,2.9){\tiny No};        
       \def\ubbot{1.2}; 
       \draw [rounded corners=2mm, ultra thick, draw=ProcessBlue,->] 
             (0,2) -- (0,\ubbot) -- (5.5,\ubbot) -- (5.5,2) (3,\ubbot) -- (3,\ubbot-0.3);         
      \node at (0,3.8)  [align=center,  draw=Red!75, fill=Red!20, diamond, aspect=3] {\tiny First file processing?};

      \draw [rounded corners=2mm, ultra thick, draw=ProcessBlue,->]
             (5.5,3.2) -- (5.5,2.4);
      \node at (5.1,2.85) {\tiny No};
      \draw [rounded corners=2mm, ultra thick, draw=ProcessBlue,->]
             (6,3.8) -- (11,3.8) -- (11,3.4);
      \draw [rounded corners=2mm, ultra thick, draw=ProcessBlue,<-]
             (7.9,2.0) -- (11,2.0) -- (11,2.6);             
      \node at (5.5,3.8)  [align=center,  draw=Red!75, fill=Red!20, diamond, aspect=3] {\tiny Is Warming Up??};
      \node at (11, 3)  [align=center,  draw=Plum!75, fill=Plum!20, text width=5 cm,  font=\tiny, rounded corners=2mm] {\scriptsize Use increased   in \verb!WUC! times iterations number of the CNLS solver algorithm};
      \node at (5.5, 2)  [align=center,  draw=PineGreen!75, fill=PineGreen!20, text width=4.0 cm, rounded corners=2mm, font=\tiny] {\scriptsize Use  \verb!x0_seed!=\verb!Channels.x0_init!};
      \node at (0,2)  [align=center,  draw=PineGreen!75, fill=PineGreen!20, text width=4.5 cm, rounded corners=2mm, font=\tiny] {\scriptsize Use the best result from the previous data file as an initial guess point \verb!x0_seed!};  
      \node at (9,4) {\tiny Yes}; 
   \end{scope}
   \begin{scope}[shift={(0.025,0.75)}]
           \draw [rounded corners=2mm, ultra thick, ->,draw=ProcessBlue] 
                 (6.3,-6) -- (6.3,-7.25) ;
                 (4.3,-6) -- (4.3,-7.25) ;   
           \draw [rounded corners=2mm, ultra thick, ->,draw=ProcessBlue, dashed] 
                 (0.3,-6) -- (0.3,-7.25) ; 
           \draw [rounded corners=2mm, ultra thick, ->,draw=ProcessBlue] 
                 (-0.7,-6) -- (-0.7,-7.25) ;                  
           \draw [rounded corners=2mm, ultra thick, ->,draw=ProcessBlue] 
                 (-1.7,-6) -- (-1.7,-7.25) ;                 
          \def\parshift{0.45}; 
          \node at (2.0+3*\parshift,-3.4+3*\parshift) [draw=Blue!75, fill=Blue!20, draw, rounded corners=2 mm, minimum 
                               width=3.0cm,minimum height=3cm,  text width=10.5cm, text height=5.5 cm]{} ;
          \node at (6.9+3*\parshift,-0.4+3*\parshift) {\footnotesize Worker \verb!N!};                               
          \node at (2.0+2*\parshift,-3.4+2*\parshift) [draw=Blue!75, fill=Blue!20, draw, rounded corners=2 mm, minimum 
                               width=3.0cm,minimum height=3cm,  text width=10.5cm, text height=5.5 cm, dashed]{};
          \node at (6.9+2*\parshift,-0.4+2*\parshift) {\footnotesize Worker ...};                               
          \node at (2.0+\parshift,-3.4+\parshift) [draw=Blue!75, fill=Blue!20, draw, rounded corners=2 mm, minimum 
                               width=3.0cm,minimum height=3cm,  text width=10.5cm, text height=5.5 cm]{};
          \node at (6.9+\parshift,-0.4+\parshift) {\footnotesize Worker 2};
          \node at (2.0,-3.4) [draw=Blue!75, fill=Blue!20, draw, rounded corners=2 mm, minimum 
                               width=3.0cm,minimum height=3cm,  text width=10.5cm, text height=5.5 cm]{};
          \node at (6.9,-0.4) {\footnotesize Worker 1};
          \draw [ultra thick,->,draw=ProcessBlue] (-1.7,-1.5)--(-1.7,-3.8);
          \draw [ultra thick,->,draw=ProcessBlue, rounded corners=2mm] (2.2,-1.5)--(2.2,-3) -- (-0.55,-3);
          \draw [ultra thick,->,draw=ProcessBlue, rounded corners=2mm] (6,-1.5)--(6,-4.3) -- (4,-4.3) ;    
          \draw [ultra thick,->,draw=ProcessBlue, rounded corners=2mm] (3,-4.3) --(1,-4.3) ;     
          \draw [ultra thick,->,draw=ProcessBlue, rounded corners=2mm] (7,-1.5)--(7.8,-1.5) -- (7.8,-5.3) -- (6.7,-5.3);    
          \draw [ultra thick,->,draw=ProcessBlue, rounded corners=2mm] (3,-5.3) --(1,-5.3) ;  
          \draw [ultra thick,draw=ProcessBlue, rounded corners=2mm] (-1.7,-4) --(-1.7,-7) ;    
          \draw [ultra thick,->,draw=ProcessBlue] (-0.2,-1.5)--(0.6,-1.5) node [midway, above] {\tiny Yes};
          \draw [ultra thick,->,draw=ProcessBlue] (-1.7,-2.5)--(-1.7,-2.7);
          \node at (-2.05,-2.45) {\tiny No};          
          \node at (-1.7,-1.5)  [align=center,  draw=Red!75, fill=Red!20, diamond, aspect=2] {\tiny Use MC mode?};
          
          \draw [ultra thick,->,draw=ProcessBlue] (3.8,-1.5)--(4.6,-1.5) node [midway, above] {\tiny No}; 
          \node at (1.8,-2.6) {\tiny Yes};  
          \node at (2.2,-1.5) [align=center,  draw=Red!75, fill=Red!20, diamond, aspect=2] {\tiny \verb!Stat_exp_num!=1};
          
          \node at (5.7,-3.2) [rotate=90] {\scriptsize \verb!local!};
          \node at (7.5,-3.2) [rotate=90] {\scriptsize \verb!global!};          
          \node at (6,-1.5) [align=center,  draw=Red!75, fill=Red!20, diamond, aspect=2] {\tiny MC mode is};
          \node at (-1.7,-3) [draw=green!50,fill=green!20]{\scriptsize \verb!x0!=\verb!x0_seed!};  
          \node at (2.6,-4.3) [draw=green!50,fill=green!20]{\scriptsize \verb!x0!=\verb!x0_seed!+$\vec{r}$};   
          \node at (4,-5.3) [draw=green!50,fill=green!20]  {\scriptsize \verb!x0!=\verb!Channels.x0_space_origin!+$\vec{r}$}; 
          \node at (-1.4,-5) [draw=Yellow!50, fill=Yellow!20, text width=4.0 cm, rounded corners = 2mm]
                             {\small Solve Eq.~\eqref{eq:Working_Function} for\\
                              $Y_m$=\verb!Channels.Model!\\ by \verb!options.method!\\ starting from \verb!x0! point};
          \node at (2.2,-7.9-0.1) [fill=Red!35, draw=Red!50, font=\tiny, text width=8cm] 
                              {\footnotesize Saving the result for minimal residual over all MC iterations to \verb!Channel.Best! and storing the statistical data to \verb!Channels! fields \verb!Best_CI!, \verb!Mean!, and \verb!CI! };
           \def\hpar{1.5}; 
           \draw [rounded corners=2mm, ultra thick, <->,draw=ProcessBlue] 
                 (6.3,0.75) -- (6.3,\hpar) -- (-1.7,\hpar) -- (-1.7,-0.6); 
           \draw [rounded corners=2mm, ultra thick, ->,draw=ProcessBlue] 
                 (3,\hpar)--(-0.7,\hpar)--(-0.7,-0.2);
           \draw [rounded corners=2mm, ultra thick, ->,draw=ProcessBlue, dashed] 
                 (3,\hpar)--(0.3,\hpar)--(0.3,0.3);                 
                
    \end{scope}
  \node at (2.2,-9.5)  [align=center,  draw=Green!75, fill=Green!20, diamond, aspect=2.5] {\scriptsize All data processed? };
  \node at (8,-9.5) [align=center,  draw=Dandelion!75, fill=Dandelion!20, rounded corners=2mm,  text width=3.5 cm, font=\tiny] {\scriptsize Saving data and notifying the operator about the end of approximation};
\end{scope}
  \begin{scope}[scale=0.9, shift={(6.5,-1.5)}] 
  \node at (5.5,1.8) {\scriptsize Evaluating  :}; 
  \node at (5.2,2.5) [draw=Gray!50] { \verb!NELM_Func.m!};
  \begin{scope}[shift={(3.5,2.6)}]
      \draw [very thick, draw=Green, rounded corners=1.5mm,->] (-0.2,0.2) -- (-0.2,-0.2)--(0.1, -0.2);      
  \end{scope}  
  \node at (5.5,-0.5) [draw=Goldenrod!50, fill=Goldenrod!20]{\tiny \begin{tabular}{p{1.7 cm}|p{1.7 cm}}
                          Parameters for minimum residual               & \verb!Channels.Best!\\ & \\
                          Parameters  CI               & \verb!Channels.Best_CI!\\ & \\
                          MC mean parameters value                & \verb!Channels.Mean!\\ & \\
                          MC CI               & \verb!Channels.CI!\\ & \\
                          Solver info output       & \verb!Channels.Info!\\
                          Error info      & \verb!Message!\\
                        \end{tabular}};
 \end{scope}   
  \begin{scope}[scale=0.9, shift={(-4.5,-2.4)}]
  \node at (0,0) [draw=Plum!50, rotate=90] {\footnotesize Parallel \verb!parfor! loop with \verb!Stat_exp_num! iterator};
 \end{scope}  
 \begin{scope}[shift={(9.5,-6.5)}]
 \node at (0,0) [draw=Gray!50] {\verb!NELM.m!};
  \begin{scope}[shift={(-2.3,-0.15)},yscale=-1,xscale=1]
      \draw [very thick, draw=Green, rounded corners=1.5mm,->] (-0.2,0.2) -- (-0.2,-0.2)--(0.1, -0.2);      
  \end{scope}
    \begin{scope}[shift={(2.5,0.25)}]
      \draw [very thick, draw=Green, rounded corners=1.5mm,->] (-0.3,-0.2)--(0.1, -0.2);      
  \end{scope}
 \end{scope}

\end{tikzpicture}
\caption{NELM Software Block Diagram. Here $\vec{r}$ is a random vector, MC is Monte-Carlo, CI is confidence intervals.}
    \label{fig:NELM_sketch}
\end{figure*}
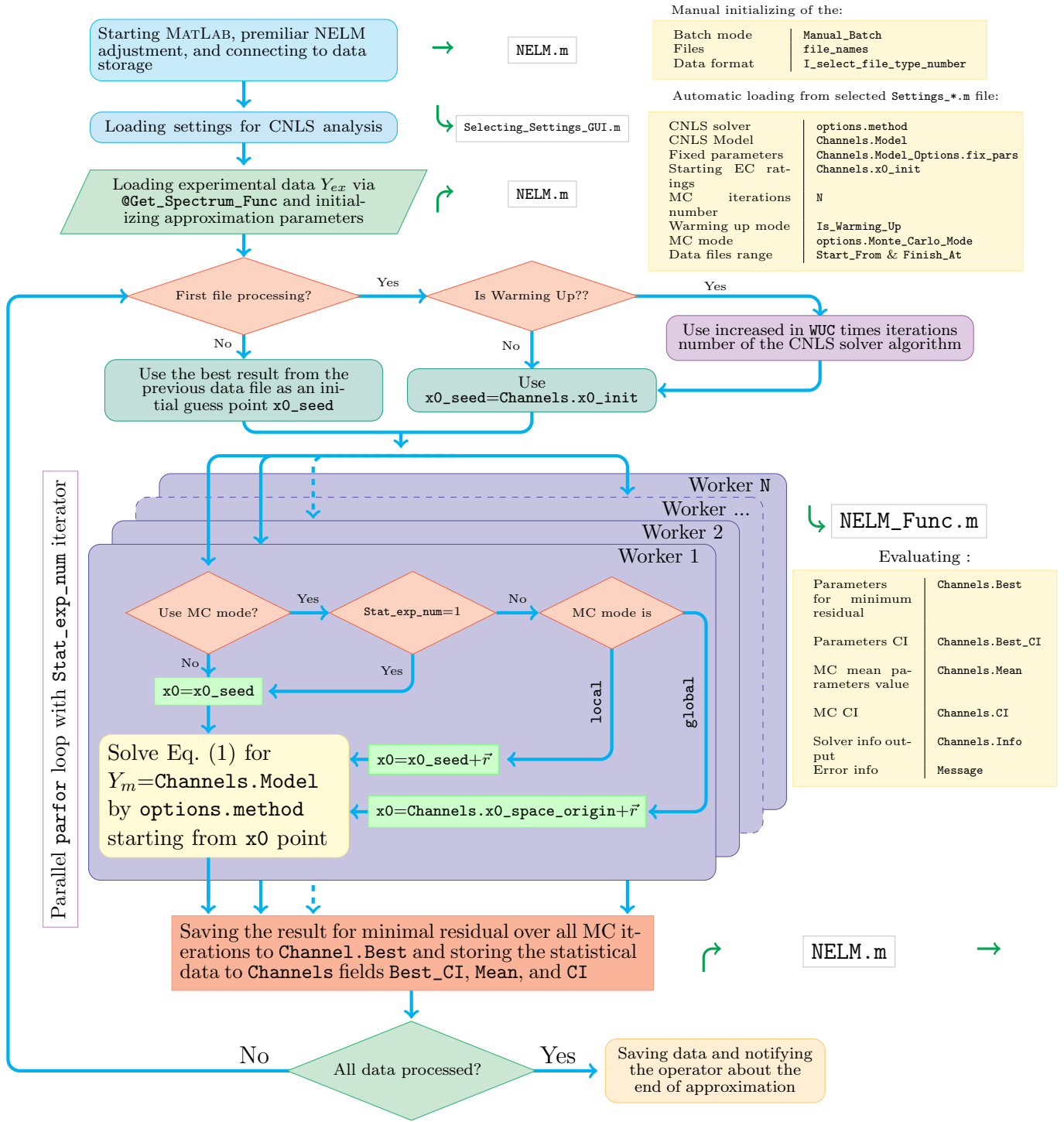
\begin{figure}
    \centering
    \subfigure[]{    \begin{tikzpicture}
        \node at (15,7) {\includegraphics[width=\textwidth]{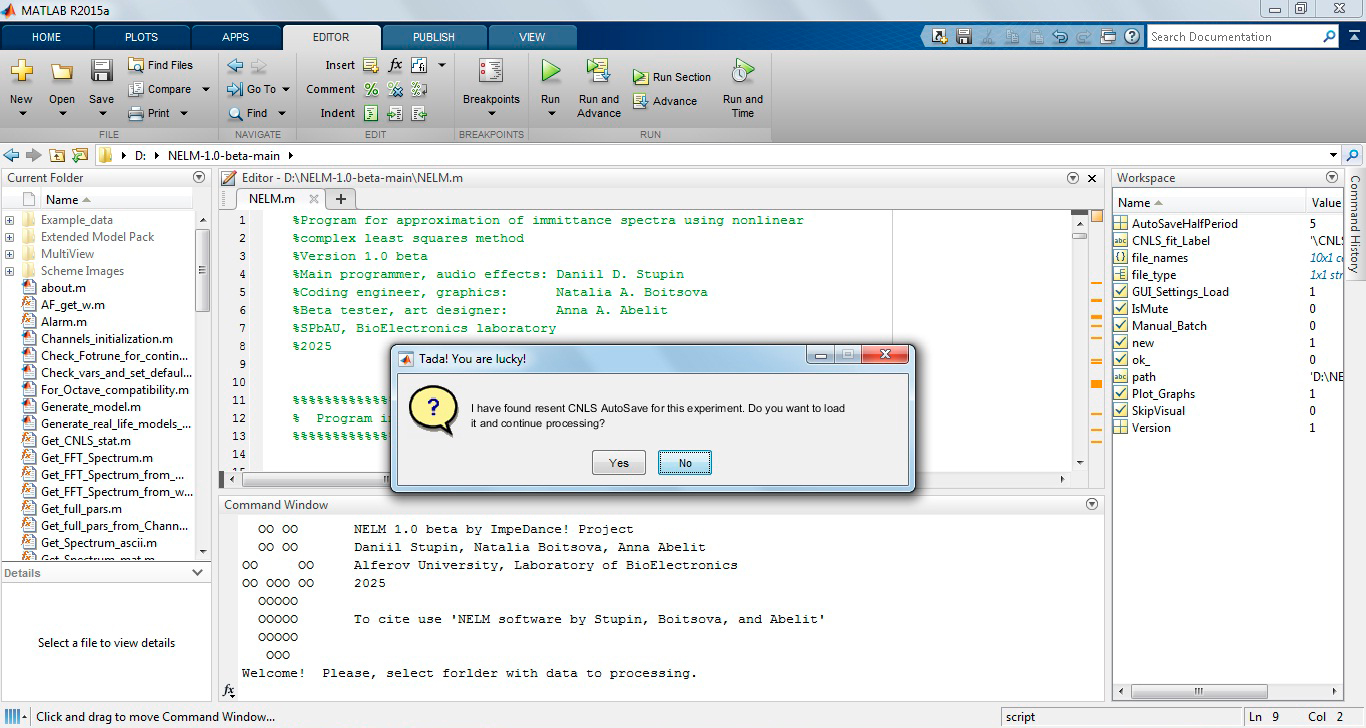}};
    \end{tikzpicture}} \\ 
    \subfigure[]{    \begin{tikzpicture}
                       \node at (12,1.8) {\includegraphics[scale=0.45]{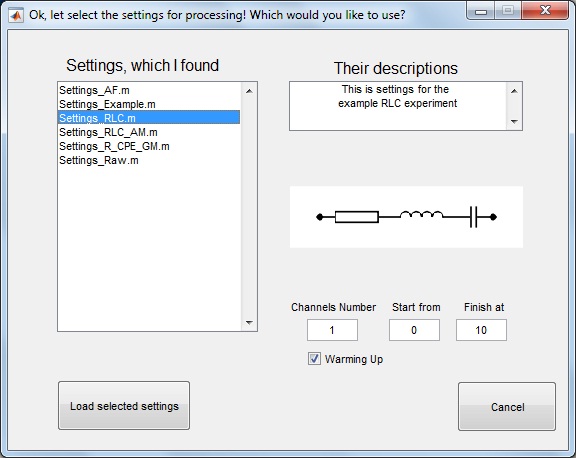}};
                      \end{tikzpicture}}

     \caption{
    Screenshots of NELM software. (a) \textsc{MatLab} editor window in which the NELM shell code is open, and a message prompting to continue the previously started CNLS processing of spectra; (b) window for selecting an equivalent circuit for data processing.
    }
    \label{fig:NELM_Screens_1}
\end{figure}
\newpage
\thispagestyle{empty}
\begin{figure}
    \centering
    \subfigure[]{    \begin{tikzpicture}
        \node at (15,7) {\includegraphics[width=\textwidth]{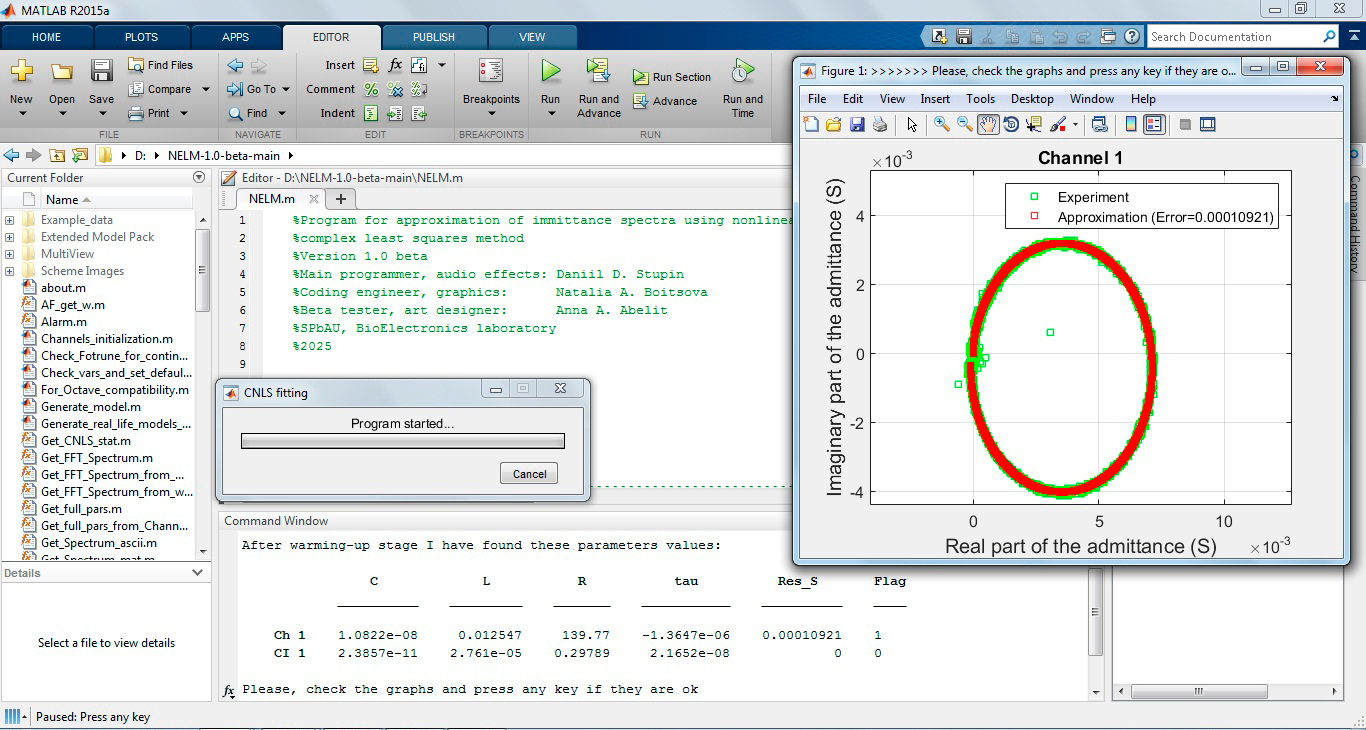}};
    \end{tikzpicture}} \\ 
    \subfigure[]{    \begin{tikzpicture}
                       \node at (12,1.8) {\includegraphics[width=\textwidth]{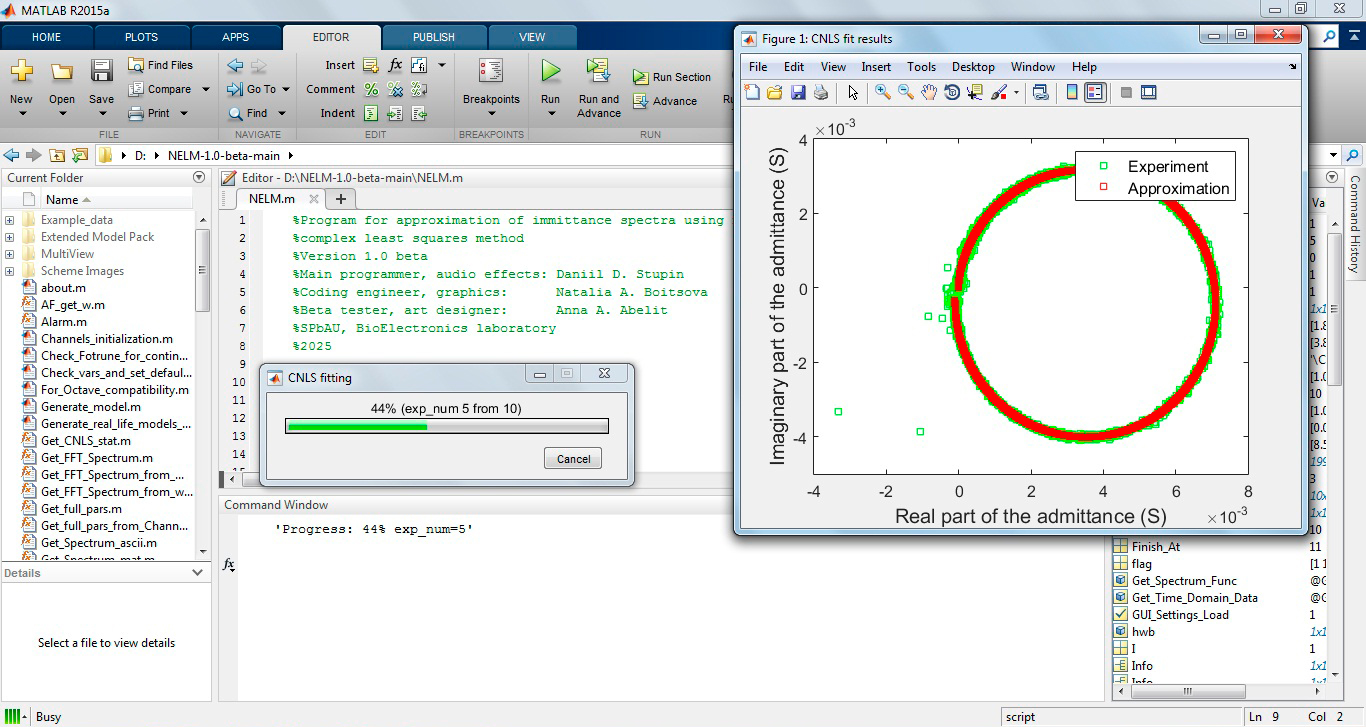}};
                      \end{tikzpicture}}
     \caption{
Screenshots of NELM software (continued). (a) \textsc{MatLab} editor window after the warm-up stage. The screenshot shows a graph with the approximation results and a table with the EC elements ratings. Here CI are confidence intervals;  (b) the result of CNLS processing of one of the RLC-admittance spectra and a window displaying the number of already processed spectra.
    }
    \label{fig:NELM_Screens_2}
\end{figure}
\FloatBarrier
\section{Step-by-step guide and program block-scheme}
In this section we will present step-by-step instructions for NELM usage and simultaneously discuss NELM code. For this purpose we will describe each step in the text and refer to the NELM block-diagram Fig.~\ref{fig:NELM_sketch} and NELM screenshots Fig.~\ref{fig:NELM_Screens_1} and Fig.~\ref{fig:NELM_Screens_2}  to show how our package operates. So, to process impedance data with NELM the following workflow should be done (for video lessons see playlist on YouTube \cite{NELM_YouTube}).
\begin{enumerate}
    \item At first, before running main script \verb!NELM.m! user  can choose the input batch processing type, data format,  and initial path, where experimental data is stored. This can be done by using variables \begin{center} \verb!Manual_Batch!, \verb!names!, \verb!file_type! \end{center} respectively in the script beginning. In addition, since impedance data, especially in time representation, may take a large storage space, the user can connect to external network (cloud) where data are stored, by using, for example, VPN services. We have tested our software with Radmin VPN v1.4.4642.1 for Windows OS (Famatech, Tortola).
    \item Then, after preliminary NELM configuration is done, the user should choose the folder with data of interest if batch mode is selected (\verb!Manual_Batch=false!), or to choose data files manually otherwise. 
    \item Further, the NELM will show the graphical dialog window where settings can be selected for data processing [Fig.~\ref{fig:NELM_Screens_1}(b)].
    \item After selecting the settings, the NELM starts to process the data (Fig.~\ref{fig:NELM_sketch}). If the warming-up mode is turned on (\verb!Is_Warming_Up=true!) in the settings file, the NELM will try to find the best initial points for the CNLS analysis by using an increased number of iterations during solving Eq.~\eqref{eq:Working_Function}, after which it will show to the user result in the form of table in the console and in the form of comparative plot in the figure windows [Fig.~\ref{fig:NELM_Screens_2}(a)], and will ask the user to check them. If obtained EC spectra match to the experimental data with low residual, and EC parameters are physically correct, then user can allow to NELM to continue processing by pressing any key.
    \item During the data analysis, the NELM will show to user progress in console, and in window with progress bar, which also contain the ``Cancel'' button, by pressing which the emergence NELM shutdown can be made [Fig.~\ref{fig:NELM_Screens_2}(b)].
    \item When all data is processed, the NELM alerts user about this event by sound alarm and by showing the message in console. 
    \item Finally, the NELM will save obtained data in the mat-file at the path with name \verb![path CNLS_fit_Label]!, where \verb!path! is a directory  with processed files and  \verb!CNLS_fit_Label! is a name for an output m-file. If string \verb!CNLS_fit_Label! started with \verb!\ ! then the output will be saved in the directory with experimental data, otherwise the output m-file will be saved in the parent directory with name, which includes the name of the folder with experimental data.
\end{enumerate}
It should be noted that if some errors during CNLS treatment are occurring, the NELM does not crash, but alerts the user about them and starts processing again with the last used data file. This protection is done to prevent the NELM crashes due to network issues, for example, VPN shutting down. If such an error occurs, after network is repaired, the NELM will continue processing. If other types of error, the user can try to fix them, or they can stop NELM by pressing ``cancel'' button. By doing that, they require NELM to save workspace in its own folder and show more information about the problem in \verb!Message! variable. For resuming CNLS processing from the last successful data sample after NELM emergence stop   user can set the variable \verb!new=false! at beginning of the \verb!NELM.m! script.

After acquiring the starter pack, to make sure that the program was installed correctly, we recommend having a test run of the program with the data and settings provided for RLC circuit (use the pre-made \verb!Settings_RLC.m! and data in folder\\ \verb!\Example_data\ascii\RLC 130 ohm 15 mH 10 nF\!, see video lesson \cite{NELM_YouTube}).

\section{Overview of the NELM's  immittance analysis techniques}
As discussed in the introduction, for impedance spectra analysis typically equivalent circuits concept is applied, and a goal of experimental impedance data analysis is estimation of the parameters of such equivalent circuits, which can be done, in particular, using CNLS approach [Eq.~\eqref{eq:Working_Function}].  For this purpose, in common cases numerical minimization algorithms are used \cite{nocedal1999numerical}.  This section is devoted to an overview of the strategies that were implemented in the NELM impedance spectra analysis, which we summarized for user convenience in Tab.~\ref{tab:methods}. 

\begin{table}
\footnotesize
    \centering
    \begin{tabular}{p{5 cm} p{8 cm}}
    \hline\hline
    Method & Description\\
    \hline\hline
      \verb!'Nelder-Mead'!   &\footnotesize Classical Nelder-Mead search \cite{Hansen,nocedal1999numerical,NM}.  \\ \hline
      \verb!'Levenberg-Marquardt'!  &\footnotesize Levenberg-Marquardt minimization \cite{Hansen,nocedal1999numerical}.\\ \hline
      \verb!'Trust_Region'!  &\footnotesize Trust region algorithm \cite{kelley1999iterative}.\\ \hline      
      \verb!'NMLM'! &\footnotesize Combination of the Nelder-Mead algorithm with  Levenberg-Marquardt method  refinement.\\ \hline
      \verb!'NMCD'! &\footnotesize Coordinate descent algorithm, which used one-dimensional Nelder-Mead method for each coordinate \cite{Hansen,nocedal1999numerical}.\\ \hline
      \verb!'NMSD'! &\footnotesize Nelder-Mead algorithm with gradient descent method  refinement.\\ \hline
      \verb!'NM_Shake'! &\footnotesize Nelder-Mead algorithm with local minimum jumping out feature (similar to \cite{basinhopping}).\\ \hline
      \verb!'NMGN'! &\footnotesize Nelder-Mead algorithm with damped Gauss-Newton method  refinement \cite{Hansen}\\ \hline
      \verb!'Y_Homotopy'! &\footnotesize Spectrum homotopy method for trying speeding up calculations.\\ \hline
      \verb!'AF'! &\footnotesize Adaptive filtering immittance approximation for noise canceling \cite{Stupin2017,stupin2021bioimpedance}.\\ \hline
      \verb!'AM'! &\footnotesize Algebraic method for calculation EC parameters \cite{Stupin2017,stupin2021bioimpedance,MacdonaldD}.\\ \hline
      \verb!'GM'! &\footnotesize Geometric method for calculation EC parameters (compare with \cite{Tsai}).\\ \hline
      \verb!'Raw'! &\footnotesize Non-treated spectrum output ($e.g.$ for Cell-Index usage) \cite{stupin2021bioimpedance,ke2011xcelligence,golke2012xcelligence}.\\
       \hline
    \end{tabular}
\cprotect    \caption{Basic methods for treating immittance spectra that are implemented in the NELM package. In addition, the NELM package include a simple Kramres-Kronig test (\verb!KK_score.m!) and distribution of relaxation times estimator (\verb!Simple_DRT.m!).}
    \label{tab:methods}
\end{table}

\subsection{State-of-the-art CNLS processing}
\label{sec:state_of_the_art_proc}
In the NELM package, there are a few different numerical CNLS-strategies  from which users can choose.  The optimization solvers based on Nelder-Mead, Trust Region and Levenberg-Marquardt algorithms were already implemented in \textsc{MatLab}, but some others, such as damped Gauss-Newton, gradient-descent search and coordinate descent search (and their combinations) were developed specifically for the NELM package. Since these methods were built from the ground up, a brief explanation of their implementation promises to be useful to NELM users.

Let us start with the damped Gauss-Newton iterative method \cite{nocedal1999numerical,Hansen}, which can be chosen in NELM by setting field \verb!options.method='NMGN'! in setting file. It is based on the replacement of model immittance $Y_m$ in the working function Eq.~\eqref{eq:Working_Function}  by its linear Taylor series approximation, which converts problem Eq.~\eqref{eq:Working_Function} into solution of the normal equations system \cite{Lawson_Eng, Rice_Eng}. The iterations for finding the parameter vector in this case look like 
\begin{equation} 
\label{eq:Damped_Gauss_Nerwton}
\vec{p}^{~(n+1)}=\vec{p}^{~(n)}-\lambda\Re\left( {J^+J} \right)^{-1} \Re \left(J^+ \vec{F}[\vec{p}^{~(n)}]\right), 
\end{equation} 
where $J=\nabla_{\vec{p}}Y_m $ -- Jacobian of the Gauss-Newton system,\footnote{In fact, this Jacobian $J$ is a matrix, which columns corresponds to  $Y_m$ gradient components, calculated at different frequencies.} 
$n$ -- iteration number, $ \vec{F}_k=Y_m(\omega_k, \vec{p}^{~(n)})-Y_{ex}(\omega_k)$ -- the residual between model and experiment [compare with~Eq.~\eqref{eq:Working_Function}], $^+$ denotes hermitian conjugation, $\lambda$ is so-called damping factor.

Its use allows us to expand the capabilities of the classical Gauss-Newton method, since varying the parameter $\lambda$ allows the algorithm to work correctly not only near the minimum point. To be specific, in our version of the damped Gauss-Newton algorithm at each new iteration $\vec{p}^{(n)}$ uses the bisection algorithm to choose the appropriate value for $\lambda$ in the range, $e.g.\,[0,1]$, which gives the lowest value to the working function in the direction defined by Eq.~\eqref{eq:Damped_Gauss_Nerwton}. Such a strategy not only includes the classical Gauss-Newton $\lambda=1$ value, but in the case if classical $\lambda=1$ leads to the growth of the working function it allows us to make a shot at changing the working function value to something that is lower than it's value at $\vec{p}^{(n)}$.
We found the application of this damped Gauss-Newton technique is very useful for refining the results of the Nelder-Mead solver, thus in the NELM package these algorithms preferred to be used in pairs, for which additionally \verb!options.Use_NM!  flag should be set to \verb!true!. In other case, the NELM will solve Eq.~\eqref{eq:Working_Function} by using the damped Gauss-Newton method only.

It is fruitful to note that during the execution of the damped Gauss-Newton method the convergence $\lambda$ to unity can be used as an indication whether the algorithm has reached the minimum point. Moreover,  the iterative process almost always converges, since at each step, the algorithm moves to the lower value of the working function.

Let us now consider the gradient descent method. Contrary to the Gauss-Newton approach, this iteration technique solves equation $\nabla E^2=0$ without using approximation for model immittance $Y_m$. The idea for iterations design in gradient descent approach is based on the false-position rule, namely
\begin{equation}
    \vec{p}^{~(n+1)}=\vec{p}^{~(n)}-\lambda \Re J^+ \vec{F}[\vec{p}^{~(n)}], 
\end{equation}

where $\lambda$ is a step, which mathematical meaning is similar to the Gauss-Newton damping factor, however, its choosing strategy is quite different. Contrary to the Gauss-Newton method, the steepest descent approach does not have any typical value for $\lambda$ like unity, and the effective change in the $\vec{p}^{~(n)}$  strongly depends on the magnitude of $J$. Thus, we have used some heuristic technique, which is based on the idea that the maximum relative change of the elements of $\vec{p}^{~(n)}$ should not be greater than 1\%. This strategy provides the convergence of the steepest descent method by blocking far away jumping from the $\vec{p}^{~(n)}$ point, which we assume is located in relative proximity to the minimum point.
  
Let's now take a look at the coordinate descent method, which in fact is the sequential  iterative application of the one-dimension minimization algorithms for each EC parameter. In NELM Nelder-Mead algorithm is used as a one-dimension solver, $i.e.$ minimization of the working function Eq.~\eqref{eq:Working_Function} are provided via automatical fixing values of all EC parameters except one (compare with Sec.~\ref{sec:model_description}), for which the Nelder-Mead approach is applied. At the next iteration, the next variable is passed to the solver, while other parameters are freezed and so on. Such approach can be used when Jacobian matrix  is ill-posed.
\FloatBarrier

Finally, it should be emphasized that NELM allows any immittance representation to be used for CNLS fitting, since the admittance terms $Y$ in the Eq.~\eqref{eq:Working_Function} can be replaced in the NELM code by an arbitrary immittance values via setting  \verb!options.data_type! field (Tab.~\ref{tab:types}) in the \verb!Settings_*.m! file.  Also, weight type $\mathcal{W}$ in  Eq.~\eqref{eq:Working_Function} can be selected by setting \verb!W_type! variable (\verb!unit!, \verb!inverse_moduluse! weights are already implemented in NELM, as is the \verb!custom! weight, which can be $e.g.$ loaded from a file).

\begin{table}[]

\footnotesize
    \centering
    \begin{tabular}{p{4 cm} p{7.5 cm} p{2 cm}}
    \hline\hline
    \verb!options.data_type! & Immittance used in Eq.~\eqref{eq:Working_Function} instead $Y$ & Domain \\
    \hline\hline
      \verb!'Y'!   &\footnotesize Admittance (default) & \footnotesize Complex \\ \hline
      \verb!'Z'!   &\footnotesize Impedance & Complex \\ \hline
      \verb!'abs_Y'!   &\footnotesize Admittance magnitude & Real \\ \hline 
      \verb!'abs_Z'!   &\footnotesize Impedance magnitude & Real \\ \hline 
      \verb!'real_Y'! or   \verb!'imag_Y'!   &\footnotesize Real or imaginary part of the admittance & Real \\ \hline      
      \verb!'real_Z'! or   \verb!'imag_Z'!   &\footnotesize Real or imaginary part of the impedance & Real \\ \hline  
      \verb!'angle_Y'!    &\footnotesize Phase of the admittance (experimental) & Real \\ \hline   
       \hline
    \end{tabular}
      \cprotect  \caption{Immittance types available in NELM for CNLS processing. Users can add their own types ($e.g.$ $\log_{10}|Y|$) to the \verb!Residual_LMS.m! script.}
    \label{tab:types}
\end{table}

\newpage
\subsection{Monte-Carlo CNLS stabilization}
\label{sec:MC}
Since CNLS is a non-linear problem, local  (false) minima points of the working function can exist. To our best knowledge, today there is no universal algorithms for finding global minimum developed, but some heuristic approaches still can be utilized for overcoming local minimum issues. One such promising technique can be realized by using Monte-Carlo concept, $e.g.$ by probing working function via starting minimization algorithms from different initial points with further choosing the result with minimal residual, so called Multistart approach \cite{About_Multistart}. The mentioned above starting points for other factors being equal can be chosen randomly in some  reasonable area in the EC parameters space. This strategy is implemented in NELM by utilization of the \verb!parfor! loop type for increasing the speed of the CNLS calculations. 
Moreover, to stabilize the operation of the NELM software for each spectrum in batch processing of the time-depend, non-stationary immittance data samples, we are not only using the best result of the previous approximation as a starting point for the next one, but also a certain set of points, obtained randomly in a small vicinity of mentioned above starting points (local MC mode) or near appropriate ``universal'' starting point (global MC mode), which does not change during approximation process. The random selection of these points was carried out using the \verb!rand! or \verb!randn! functions of the \textsc{MatLab} software, and their number in total was usually equal to the number of processors on the computer, which made possible for us to speed up the Monte-Carlo calculation process due to parallel processing of data carried out using the \verb!parfor! loop.
In this case, for each spectrum, the best result of the Monte-Carlo experiment (with the smallest error), as well as the average value of the parameters of the equivalent circuit and their confidence intervals obtained with its help, were stored in the computer memory in the \verb!Best!, \verb!Mean!, and \verb!CI! fields of the \verb!Channels! structure, respectively.  
This approach allows us to discard the results of the CNLS approximation with a large error and thereby solve the above-described problem of destabilization of the approximation process due to the appearance of unsuccessful CNLS fits. The obtained statistical data can be thought of as a computationally inexpensive measure of the working function surface smoothness, which can be used for measurements quality control. It should be noted, that we does not use build into \textsc{MatLab} method \verb!MultiStart! since NELM was developed as multipurpose investigating CNLS platform and  thus we try to make code flexibile for users challenging problems, for example for statistical analysis of the working function mentioned above. The on/off switching of MC mode is provided via setting the value of the  \verb!N! variable in the \verb!Settings_*.m! file, which is equal to the total number of the MC iterations: if  \verb!N!=1, then MC mode is switched off, otherwise the MC mode is turned on. The type of the MC mode (local or global) is achieved by setting field \verb!options.Monte_Carlo_Mode! to \verb!'local'! or \verb!'global'!, respectively.

Another Monte Carlo application for local minimum fighting could be done by addition to a solver's result $\vec{p}$ by some random vector $\vec{r}$ and by starting solver again using $\vec{p}+\vec{r}$ as initial point with the hope of jumping out from the possible local minimum. This strategy is implemented in \verb!NM_Shake! approach in NELM.

NELM also supports estimation of confidence intervals for CNLS fitting that is not based on Monte-Carlo trials, but on the Jacobian usage (Sec.~\ref{sec:state_of_the_art_proc}) instead. Namely, after the fitting procedure, NELM provides a working function \eqref{eq:Working_Function} Taylor linearization and applies the standard Student's approach \cite{hanson1978confidence, Ivchenko} to the resulting system of linear equations to obtain confidence intervals, which are stored in \verb!Channels.Best_CI! field.

\subsection{Adaptive filtering approach}
The NELM software development was motivated by the absence of immittance analysis programs, which were able to work with high resolution spectra, generated mainly by FFT impedance spectroscopy \cite{DFT,DFT2,Review}. Unfortunately, the FFT approach is not robust to noise and inferences contrary to the AI-based adaptive filtering approach, which also produces high-resolution spectra \cite{Stupin2017,stupin2021bioimpedance,stupin2018cell}. In order to give to users the  possibility to work with noisy data, we have also integrate into the NELM the adaptive filtering processing. Using a slight modification of the convolution theorem \cite{Brigham} we also increase the AF-based approach speed, making it comparable to the FFT approach. To call the adaptive filtering method, the user should set in the field \verb!options.method='AF'!, and declare the filtering order in the fields \verb!options.ell_n! (for voltage) and \verb!options.ell_d! (for current). 

\subsection{User-defined CNLS solvers and models}
\subsubsection{Adding new methods of minimization to NELM}
One of the key features of the NELM package is the possibility to add and modify data processing protocols. Specifically, to create a new immittance analyzing method, the user can create in \verb!NELM_Func.m! subroutine the new case in the \verb!switch options.method! environment, and then select it for data processing by changing \verb!options.method! field the \verb!Settings_*.m! file. The data processing code can be arbitrary, but should return the immittance treatment results in the \verb!Result! row-array.  Also, users' data processing code should return in the \verb!Time! variable the time, when experimental data were captured, in the \verb!info! structure it should return the information about data analysis performance (fields \verb!iterations!, \verb!funcCount!, \verb!algorithm!, and \verb!message!), in the  \verb!flag! user's code should return the machine understandable numeric pseudonym for success or unsuccessfulness of the data processing treatment, and also user's code should return \verb!Misc! structure with additional information, $e.g.$ it can contain the level of the offset voltage, temperature, $etc$. The values of the   \verb!flag! and \verb!Time!  variables and \verb!info! and \verb!Misc! fields are not essential for NELM operating, and in the simplest case their values can be set to zeros or empty strings as some kind of stub (compare with case \verb!'AF'! in the \verb!NELM_Func.m!).  In the new case for data processing, the user can load the experimental data by using anonymous function \verb!Get_Spectrum_Func!, as well as they can access the model for data fitting by calling anonymous function \verb!Model! (which is a copy of \verb!Channels.Model! in the \verb!NELM_func! environment). Both of these anonymous routines should be specified in the \verb!Setting_*.m! file. For CNLS processing in the NELM there exists routine \verb!Residual_LMS!, which calculates the least mean squares (LMS) error between a experimental and modeling data. This routine via wrapper, typically denoted in NELM as \verb!Target!,  can be sent to any LMS solver, as it is done in the current NELM version for \verb!'Nelder-Mead'! or \verb!'Levenberg-Marquardt'! methods.
All already existing cases have comments and can be used as guidance in order to create new methods of approximation.
\subsubsection{Model interface description}
\label{sec:model_description}
The NELM package version 1.0 beta already contains mostly used in physics, chemistry, and biology EC, which are listed in Tab.~\ref{tab:models}. 
For software flexibility, each model in NELM is represented as a distinct overload m-function file, with mask \verb!M_*.m!, the output mode of which is controlled by its input format. For obtaining as output the model admittance values at specified frequencies the model m-function should be called with three arguments, namely the frequencies vector, at which immittance should be calculated, the parameters of the EC, which can be tuned for fitting process, and \verb!Model_Options! structure, which controls the immittance evaluation process.   In this case, of three arguments input mode if flag-field \verb!Model_Options.get_J! is equal to true, then the model m-function will return two values: the admittance  and it's exact Jacobian calculated at frequencies input vector. Here we emphasize that by default the model functions should return the admittance and its Jacobian. For CNLS processing, Jacobian calculations, and plotting with other immittance types, the user can select the required type from Tab.~\ref{tab:types} using \verb!options.data_type! variable in \verb!Settings_*.m! file.  Along with the obvious application of the Jacobian evaluation feature for the minimization of the working function Eq.~\eqref{eq:Working_Function}, its purpose also includes the analysis of the conditionality of the CNLS problem and the search for redundant parameters of the equivalent circuit. If the model m-function is called with no arguments or with one argument \verb!Model_Options! it returns the information about the model in the form of structure, the common fields of which are listed in Tab.~\ref{tab:info_model}. The last output mode is very essential for NELM execution, since model information is frequently used in the main script, $e.g.$ for fitted parameters parsing.

The immittance evaluation inside the model m-function can be performed using all \textsc{MatLab} computation power, including pipeline calculations (\verb!.*, ./, .^! notations). In this context, it should be stressed that NELM is not limited to parallel and series connection of the circuits, but is also able to work with bridge-like circuits, which is known for its inability to be represented by these types of connection. The simplest example of such circuit types was added to the NELM package as the \verb!M_Bridge_rl.m! model (see Tab.~\ref{tab:models}). 

Moreover, the number of the EC presented in Tab.~\ref{tab:models} is much higher than its row number, since we also have taken on board the EC constructing technique from the original LEVM program.
More specifically, the user can fix certain EC parameters at preferred values by using \verb!Model_Options.fix_pars! 2D-array field, therefore, they will create a new EC. To achieve this, the user should in the first line of the \verb!fix_pars! array specify the indexes of the parameters that should be fixed, and in the second line they should type the corresponding parameter values. For example, option \begin{center}\verb!Model_Options.fix_pars=[1, 7; 8 11]!\end{center} will fix the EC first parameter value at 8, and the seventh EC parameter value at 11. Also for converting some elements into the break-circuit or short-circuit the user can set their impedance to infinite or zero, respectively, by choosing appropriate elements parameters values. For example, for creating parallel RC-circuit EC, the user can take as the basis the \verb!M_Fricke_Morse_rl.m! model, and set its CPE exponent value to 1, and set its series resistor rating to zero.
It should be stressed that after parameters values are fixed in the input initial parameters vector \verb!x0_init! the values that correspond to fixed parameters should be deleted.

\begin{table}[]
\small
    \centering
    \begin{tabular}{p{5 cm} p{8 cm}}
    \hline
    Field& Description \\ \hline
     \hline
    \verb!Elements_names! & The cells-array, containing the names of the EC parameters. \\ \hline
    \verb!Pars_N! & Total number of the EC parameters. \\ \hline
    \verb!Elements_Num! & Number of the non-fixed EC parameters. \\ \hline  
    \verb!non_fix_pars_idx! & The indexes of the non-fixed parameters. \\ \hline
    \verb!non_negative_pars! & The indexes of the non-negative parameters with respect to the non-fixed parameters order.\\ \hline
    \end{tabular}
    \caption{The fields of the information structure, which return NELM impedance-model routines, when they are called with one argument. 
    Theese info-structures are essential for NELM functioning.}
    \label{tab:info_model}
\end{table}
Along with the LEVM concept for constructing an EC, the user can define their own EC by using the \verb!Make_Model! routine.  
In order to use it, users only need to pass into this function admittance of the new model as symbolic expression (\verb!syms! type in \textsc{MatLab}), name of the variable, which corresponds to frequency (in format $i \omega$, $e.i.$ it must be a single symbolic expression denoting the product of the imaginary unit and the angular frequency), file name of EC model m-function, and names of variables that can become negative. For example, to define EC for the single CPE element model, the user should calculate its admittance manually ($e.i.$ \verb!Y=W*(iw)^alpha!) and call function
\begin{center} \verb!Make_Model(Y, iw, 'CPE', alpha,0)! \end{center}  with additional parameters \verb!iw! (frequency name), \verb!'CPE'! (function file name) and \verb!alpha! (parameters that can drop below zero), after that \textsc{MatLab} will automatically create CPE.m function file, containing both EC admittance calculation part and  Jacobian calculation part implemented in m-code. Here we want to stress that \verb!Make_Model! function for Jacobian evaluation uses \textsc{MatLab} low-level code, which is obtained by converting the Jacobian symbolic expression to algebraic \textsc{MatLab} pipeline routines. This allows the user to create Jacobian automatically and significantly increase the speed of Jacobian calculations.

\FloatBarrier
\newpage
\thispagestyle{empty}
\begin{table}[]
\footnotesize
    \centering
    \begin{tabular}{C{3 cm}C{4 cm}p{4 cm}p{4 cm}}
    \hline \hline
    m-file & Circuit &Description& Applications \\
    \hline \hline
\verb!M_RLC_rl.m!&     
\resizebox{4 cm}{!}{
\begin{tikzpicture}[scale=1,circuit ee IEC, thick]
 \draw (-1.5,0) to[contact] (-1.5,0) -- (-1.4,0) to[resistor]  (0.1,0) to[inductor] (1.6,0) to[capacitor] (2.5,0) to[contact] (2.5,0);
\end{tikzpicture}}& \footnotesize Simple series RLC-circuit. & \scriptsize Resonance systems ($e.g.$ quartz resonators), capacitive systems (metal/electrolyte interfaces, pn-junctions) with parasitic inductance. \\
     \hline
\verb!M_RL_CPE_rl.m!&       
\resizebox{4 cm}{!}{
\begin{tikzpicture}[scale=1,circuit ee IEC, thick]
\draw (1.6,0)  -- (1.45+1.55,0);
 \draw (-1.5,0) to[contact] (-1.5,0) -- (-1.4,0) to[resistor]  (0.1,0) to[inductor] (1.6,0)  node {\CPE} (1.45+1.55,0) -- (1.6+1.55+0.2,0) to[contact] (1.6+1.55+0.2,0);
\end{tikzpicture}} & \footnotesize Series RL-CPE circuit. & \scriptsize Electrochemical or non-ideal interfaces with parasitic inductance. \\
     \hline  
 \verb!M_R_CPE_rl.m!&       
\resizebox{3 cm}{!}{
\begin{tikzpicture}[scale=1,circuit ee IEC, thick]
\draw (0.1,0)  -- (1.5,0);
 \draw (-1.7,0) to[contact] (-1.7,0) -- (-1.4,0) to[resistor]   (0.1,0)  node {\CPE} (1.5,0) -- (2,0) to[contact] (2,0) ;
\end{tikzpicture}}& \footnotesize Series R-CPE circuit. & \scriptsize Diffusion, rough,  non-ideal electrochemical electrodes, complex distributed systems ($e.g.$ tissues). \\
     \hline   
\verb!M_Fricke_Morse_rl.m!&       
\resizebox{3 cm}{!}{
\begin{tikzpicture}[scale=1,circuit ee IEC, thick]
\draw (0.1,0)  -- (1.5,0);
 \draw (-1.7,0) to[contact] (-1.7,0) -- (-1.4,0) to[resistor]   (0.1,0)  node {\CPE} (1.5,0) -- (2,0) to[contact] (2,0) ;
 \draw (-1.3,0) -- (-1.3,0.8)  to[resistor] (1.6,0.8)--(1.6,0); 
 \node at (0,1.2) {~};
\end{tikzpicture}}& \footnotesize Series R-CPE circuit with resistive leakage also known as Cole-Cole model or Fricke and Morse circuit & \scriptsize Tissues and cells suspensions, electrochemical systems with Faradic currents, systems with resistive parasitic leakage. \\
     \hline     
\verb!M_Randles_rl.m!&       
\resizebox{3 cm}{!}{
\begin{tikzpicture}[scale=1,circuit ee IEC, thick]
\draw (0.1,0)  -- (1.5,0);
 \draw (-3,0) to[contact] (-3,0) to[resistor]  (-1.4,0) to[resistor]   (0.1,0)  node {\CPE} (1.5,0) -- (2,0) to[contact] (2,0) ;
 \draw (-1.3,0) -- (-1.3,0.8)  to[capacitor] (1.6,0.8)--(1.6,0); 
 \node at (0,1.2) {~};
\end{tikzpicture}}& \footnotesize Often used Randles circuit  & \scriptsize Electrochemical interfaces.  \\
     \hline     
 \verb!M_Cole_Cole_FM_rl.m!&       
\resizebox{3 cm}{!}{
\begin{tikzpicture}[scale=1,circuit ee IEC, thick]
\draw(0.1,0)--(1.5,0);
  \draw (-1.7,0) to[contact] (-1.7,0) -- (-1.4,0) to[resistor]   (0.1,0)  node {\CPE} (1.5,0) -- (2,0) to[contact] (2,0) ;
 \draw (-1.3,0) -- (-1.3,0.8)  to[resistor] (1.6,0.8)--(1.6,0); 
 \node at (0,1.2) {~};
 \node at (1,-0.8) {$Y\sim(i\omega\tau)^\alpha$};
\end{tikzpicture}}& \footnotesize Fricke and Morse topology scheme with Cole-Cole $\tau$-representation for CPE  & \scriptsize Measuring  time-constant for  tissue  or electrochemical interfaces.  \\ 
     \hline     
\verb!M_Lapicque_rl.m!&       
\resizebox{3 cm}{!}{
\begin{tikzpicture}[scale=1,circuit ee IEC, thick]
\draw (-0.8,0)  -- (0.6,0);
 \draw (-3,0) to[contact] (-3,0) to[resistor]  (-0.8,0) node {\CPE}   (0.6,0) -- (1.5,0) to[contact] (1.5,0) ;
 \draw (-1.1,0) -- (-1.1,0.8)  to[resistor] (1,0.8)--(1,0); 
 \node at (0,1.2) {~};
\end{tikzpicture}}& \footnotesize Lapicque topology scheme with CPE  & \scriptsize Electrochemical interfaces or pn-junctions with defects, and tissue studying. \\

\hline  
\verb!M_Multiple_R_CPE_rl.m!&       
\resizebox{3 cm}{!}{
\begin{tikzpicture}[scale=1,circuit ee IEC, thick]
\draw (0.1,0) -- (1.4,0);

\draw (0.1,-1.6) -- (1.4,-1.6);

\draw (0.1, 0.8) -- (1.4,0.8);

\draw (-3,-0.4) to[contact] (-3,-0.4) to[inductor]  (-1.3,-0.4);
 \draw  (-1.3,0) to[resistor]   (0.1,0)  node {\CPE} (1.4,0) -- (1.8,0) ;
 \draw (1.8,-0.4) --(2.2, -0.4)  to[contact] (2.2,-0.4) ;
 \draw (-1.3,0) -- (-1.3,0.8)  to[resistor]   (0.1,0.8)  node {\CPE} (1.4,0.8) -- (1.8,0.8) -- (1.8,0); 
 \draw (-1.3,0) -- (-1.3,-1.6)  to[resistor]   (0.1,-1.6)  node {\CPE} (1.4,-1.6) -- (1.8,-1.6) -- (1.8,0);
 \draw (-1.3,-0.8) -- (-0.4, -0.8);
 \draw (0.4, -0.8) -- (1.8,-0.8);
 \node at (0, -0.8) {\Huge ...};
 \node at (0,1.2) {~};
\end{tikzpicture}}& \footnotesize Parallel connection of the series R-CPE circuits. & \scriptsize 	 Unhomogeneous  or hierarchical systems ($e.g.$ embedded into electrolyte electrode with two or more surface regions with different properties).\\
\hline     
\verb!M_Bridge_rl.m!&       
\resizebox{3 cm}{!}{
\begin{tikzpicture}[scale=1,circuit ee IEC, thick]
 \draw (-3,0) to[contact] (-3,0) --  (-2,0) to[resistor]   (0, -2) -- (0, -2)-- (1, -1) node [draw=black, fill = white, thick,rotate = 45] {CPE} (1, -1) -- (2, 0) -- (3,0) to[contact] (3,0) ;
  \draw (-2,0) to[resistor]   (0, 2) -- (0, 2)-- (1, 1) node [draw=black, fill = white, thick,rotate = -45] {CPE} (1, 1) -- (2, 0) ;
  \draw (0,2) to[resistor] (0,-2) -- (0, -2);
   \draw[white, thick] (0, -2) -- (1, -3);
 \draw[white, thick] (0, 2) -- (1, 3);

\end{tikzpicture}}& \footnotesize Famous circuit, which does not reduce to parallel and series connections  & \scriptsize Electrochemical interfaces with two different regions, between which current can flow. \\
\hline
\verb!M_AF.m!&       
\resizebox{3 cm}{!}{
\begin{tikzpicture}[scale=1,circuit ee IEC, thick]
 \draw (-2,0) -- (2 ,0);
 \draw (-2,1) -- (0 ,0);
 \draw (-2,-1) -- (0 ,0);
 \draw[color=black, fill = white] (0,0) circle [radius=1];
  \node (0, 0) {\Huge $\sum$};
\end{tikzpicture}}& \footnotesize Artificial intelligence adaptive filtering model  & \scriptsize Systems which are investigated under noise environment $e.g.$ biosensors.  \\
     \hline           
    \end{tabular}
   \cprotect \caption{ Build-in NELM models presented in version 1.0 beta. Although it may seems that some of the models, for example, \verb!M_Fricke_Morse_rl.m! and \verb!M_Cole_Cole_FM_rl.m!, are nearly identical,  they both were added to spare some user's time and to avoid numerical issues which appear with different Jacobian computation. The \verb!'_rl'! (real life) suffix is used for models that include series parasitic inductance \cite{Stupin2017} (not shown in the pictures) and current-voltage delay  (see Sec.~\ref{sec:Intro}).}

    \label{tab:models}
\end{table}
\FloatBarrier
\section{Software design materials and methods}
\subsection{Hardware and software requirements}
The NELM-software was created in \textsc{MatLab} language using the Open-Source concept and for this reason it was provided with detailed comments. In the process of developing NELM software, we used libraries built into \textsc{MatLab}, $i.e.$ routines created by third party developers were not involved.
Graphical interfaces were created with the support of the \verb!guide! constructor built into \textsc{MatLab}. 

Primarily, the software was tested on three computers with the following characteristics: an Asus laptop with an Intel Corei3 2330M 2.2 GHz processor, 2 cores, 3 GB of RAM (Asus, Taiwan), a Huawei laptop with a 12th Gen Intel(R) Core(TM) i5-1240P processor 1.70 GHz, 12 cores, 16 GB RAM (Huawei, China),
Key desktop computer with Core i5-4570 3.2 GHz processor, 4 cores, 4 GB (Key, Russia). The highest performance was achieved on a Huawei laptop. 

Before starting work with NELM, users need to make sure that their computer meets all necessary requirements. To operate, NELM software requires a preinstalled application package \textsc{MatLab} version not lower than 2015a or \textsc{GNU Octave} version no lower than 9.3.0. The NELM 1.0 beta distribution requires 37 MB of free disk space (2 MB without example data). To fully use NELM, it is desirable to have a multicore processor compatible with the Parallel ToolBox package, at least 1 GB of free RAM, and access to the Internet for cloud computing. 

If the user's device meets all the criteria mentioned above, they need to download the NELM from GitHub using link \cite{NELM_download}. 

\subsection{Structure of the program and operational principle}

General steps, on which NELM software approximation is based, are presented (Fig.~\ref{fig:NELM_sketch}). Firstly, the user is required to select the folder with input data files if flag \verb!Manual_Batch! is set to \verb!true! or select manually the files of interest if flag \verb!Manual_Batch! is set to \verb!false!.

Secondly, the operator, using a graphical interface, selects a settings file, which contains information about the method of data processing, frequency and time ranges, the model for approximation and the initial values of parameters, parameters that can change during the approximation process, and ones that should have fixed values, $etc$.

After user-controlled initialization of the NELM software settings is completed, the computer begins batch CNLS processing of the spectra. Since the best initial approximation for the parameters of the equivalent circuit is not always precisely known, the program can optionally begin with a warming-up mode (if flag \verb!Is_Warming_Up!=1), the purpose of which is to increase the number of iterations by a factor \verb!WUC! to help in more accurately finding the values of the elements of the equivalent circuit for the first measured spectrum. For the remaining spectra, the initial point of approximation can be chosen based on the assumption that during the time between measurements of neighboring spectra the properties of the system under study change insignificantly. This allows us to use the result of the approximation of the $N-1$ spectrum as an initial approximation for the $N$ spectrum. However, at the same time, one should not lose sight of the existence of a situation in which the approximation on the $N-1$ spectrum turns out to be unsuccessful, since in this case, approximations for all subsequent spectra may also fail with a high probability. In order to deal with this problem the \verb!'global'! mode for Monte-Carlo multistart can be chosen (see \cite{Boitsova2024MC} and Sec.~\ref{sec:MC}).

After successful completion of the approximation process, the program saves the results in the form of a mat-file and notifies the operator about end of its work through the console and a sound signal.

In addition to the scripts described above, the NELM package includes a number of routines that facilitate the processing and analysis of spectra. In particular, such utilities include the \verb!MultiViewGUI! script, which allows the user to visualize both the experimentally measured spectra themselves and their CNLS approximation, script \verb!KK_score.m! for performing the Kramers-Kronig test \cite{hu1989kramers, urquidi1990applications},  and a \verb!simplde_DRT.m! --  implementation of the transformation of immittance into a relaxation time distribution \cite{gavrilyuk2017use}.

\section{NELM application for real-life systems research}
One of the important steps for any software in the final stage of development is to check its performance on the several test input examples, for which the supposed outcome is known. 
Thus, this section is devoted to demonstration and verification of NELM performance and efficiency to solve CNLS problems in various scientific areas. We have started such a demonstration with simple RLC-circuits analysis, further we will show the applicability of NELM for semiconductor device diagnostics, after that we will present a bright example of the NELM application for biosensing and electrchemistry, and finally we will demonstrate the built-in NELM AI-based noise reduction technique, namely the adaptive filtering impedance spectroscopy approach.

The samples, whose impedance spectra was measured by the setup, discussed in Refs.~\cite{Stupin2017,stupin2021bioimpedance}, were used as input for NELM. The data were collected with usage of the sweep-shaped 15 mV excitation voltage with frequency band 10 Hz -- 40 kHz.  The sampling rate used was 500 kHz, and the spectra were measured with 2 Hz resolution. For calculating high-resolution spectra, the FFT and AF approaches were used \cite{DFT, DFT2, Review, Stupin2017,stupin2021bioimpedance}.
For data analysis, we have used the admittance representation for immittance, and no weights function was used, $i.e.$ the Eq.~\eqref{eq:Working_Function} with $\mathcal{W}=1$ was solved for CNLS processing.
After collecting impedance measurement, we analyzed the obtained results with NELM using the Nelder-Mead minimalisation method and the Monte-Carlo multistart stabilization mode \cite{About_Multistart}. For all experiments, except AF-noise immunity test, as a model the RL-CPE equivalent circuit with delay exponent was used, which impedance is
\begin{equation}
\label{eq:model}
    Z=\left(R+i \omega L+\dfrac{1}{W(i \omega)^\alpha}\right)\times e^{i \omega \tau},
\end{equation}
where $R$ is resistance, $L$ is  inductance (modeling inductor or Op-Amp), $W$ is pseudocapacitance, $\alpha$ is non-ideality parameter (CPE exponent), $\tau$ is delay between voltage and current collection in ADC. 
For AF-noise immunity test we have used Lapicque model (Tab.~\ref{tab:models}) with additional inductor and delay exponent.

In order to test abilities of NELM on lumped schemes, we gathered impedance spectra from RLC circuits composed of inductor (Bouruns, UK) with factory ratings $L=$15 mH and $R$=100 $\Omega$ and capacitors with factory ratings 10, 15, and 20 nF. For demonstrating NELM applicability for semiconductor studies we have selected the silicon photodiode BPW 20 RF (Vishay, USA) and measure it's capacitance-voltage response in the absence of photocurrent. Next, we have shown the power of the NELM for biosensing and electrochemistry; to be specific, we have demonstrated single HeLa cell detection using CNLS and multielectrode array 60MEA200/30iR-ITO from MultiChannels Systems (Germany). Finally, for AF noise immnuity test, we have made Lapicque circuit with series resistance 212~$\Omega$, leakage resistance 992~$\Omega$, and capaciance 48~nF. The ratios of the values of this chain were chosen similar to the ratios of the elements rating for living cell EC.

\begin{figure*}
    \centering
    \vspace{-2 cm}
    \subfigure[]{\includegraphics[width=7 cm]{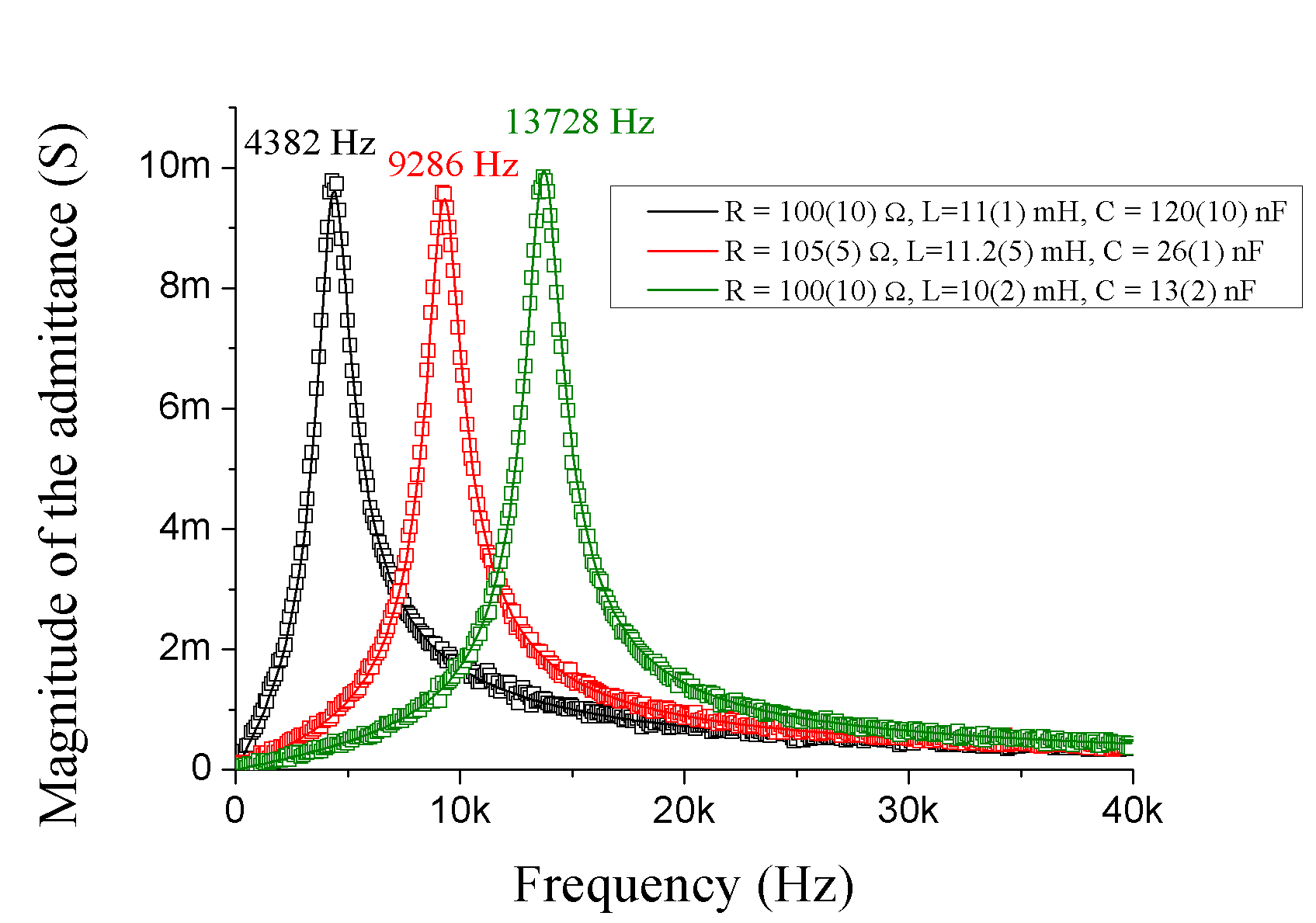}}~\subfigure[]{\includegraphics[width=7 cm]{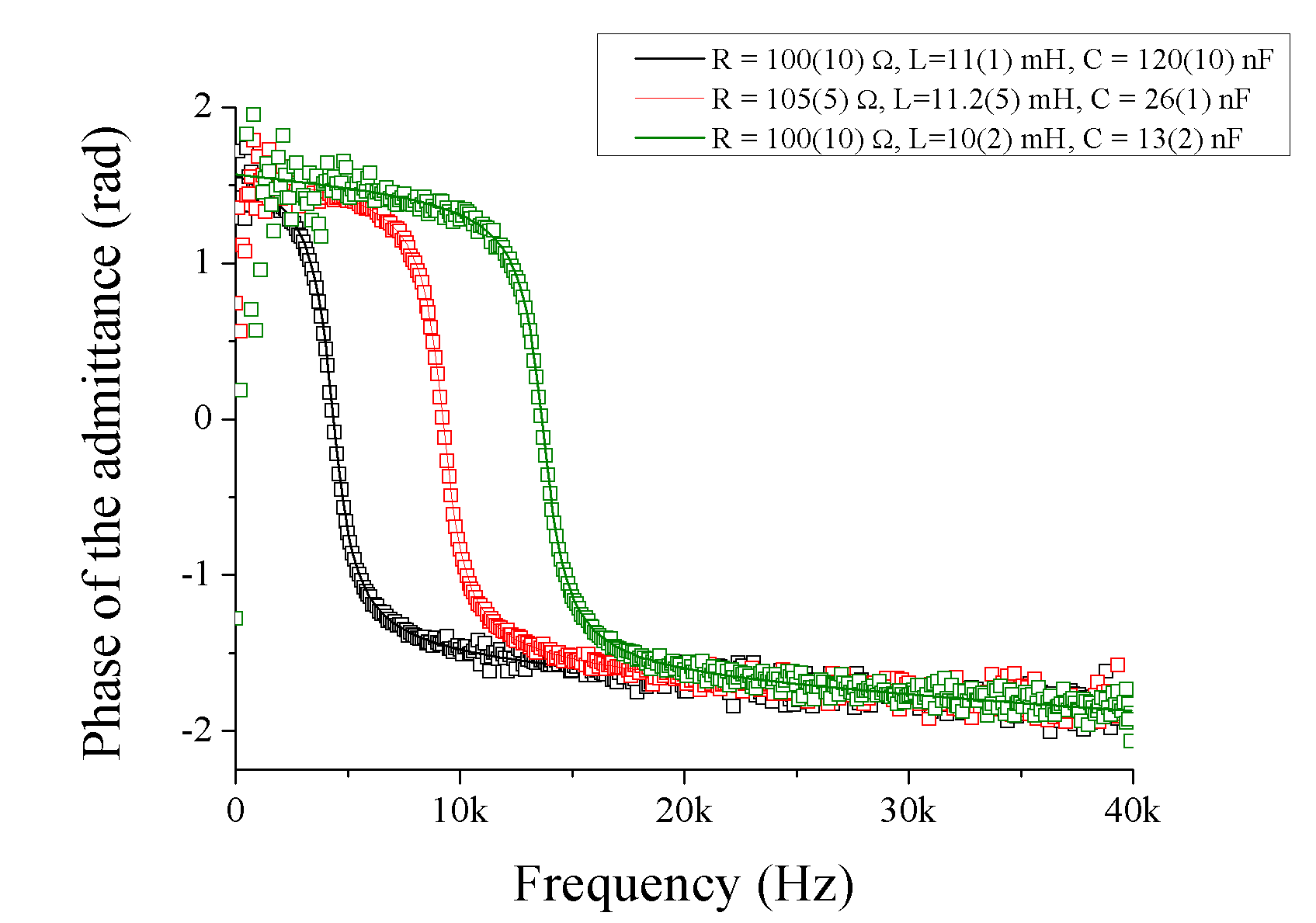}}\\ \subfigure[]{\includegraphics[width=7 cm]{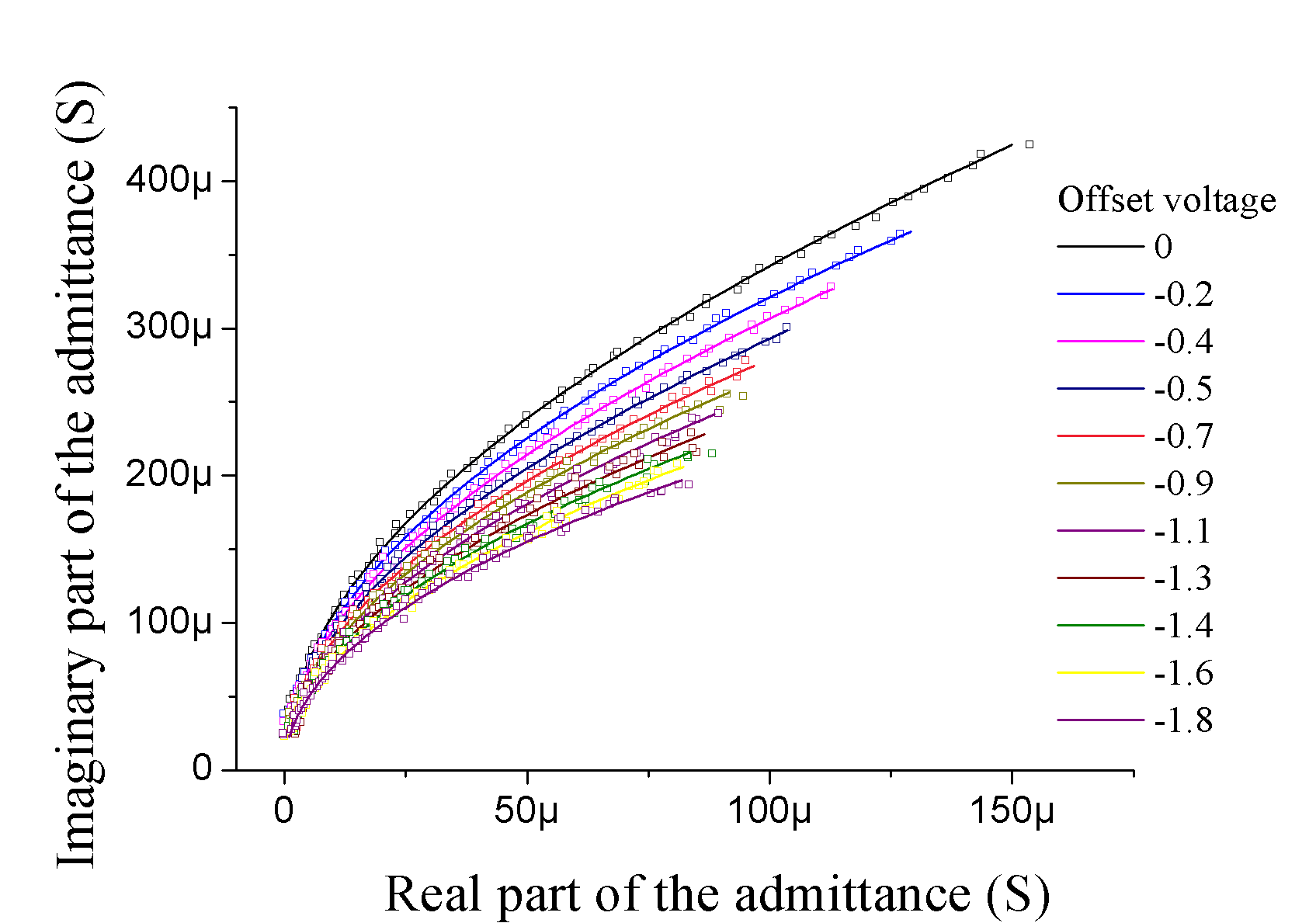}}~\subfigure[]{\includegraphics[width=7 cm]{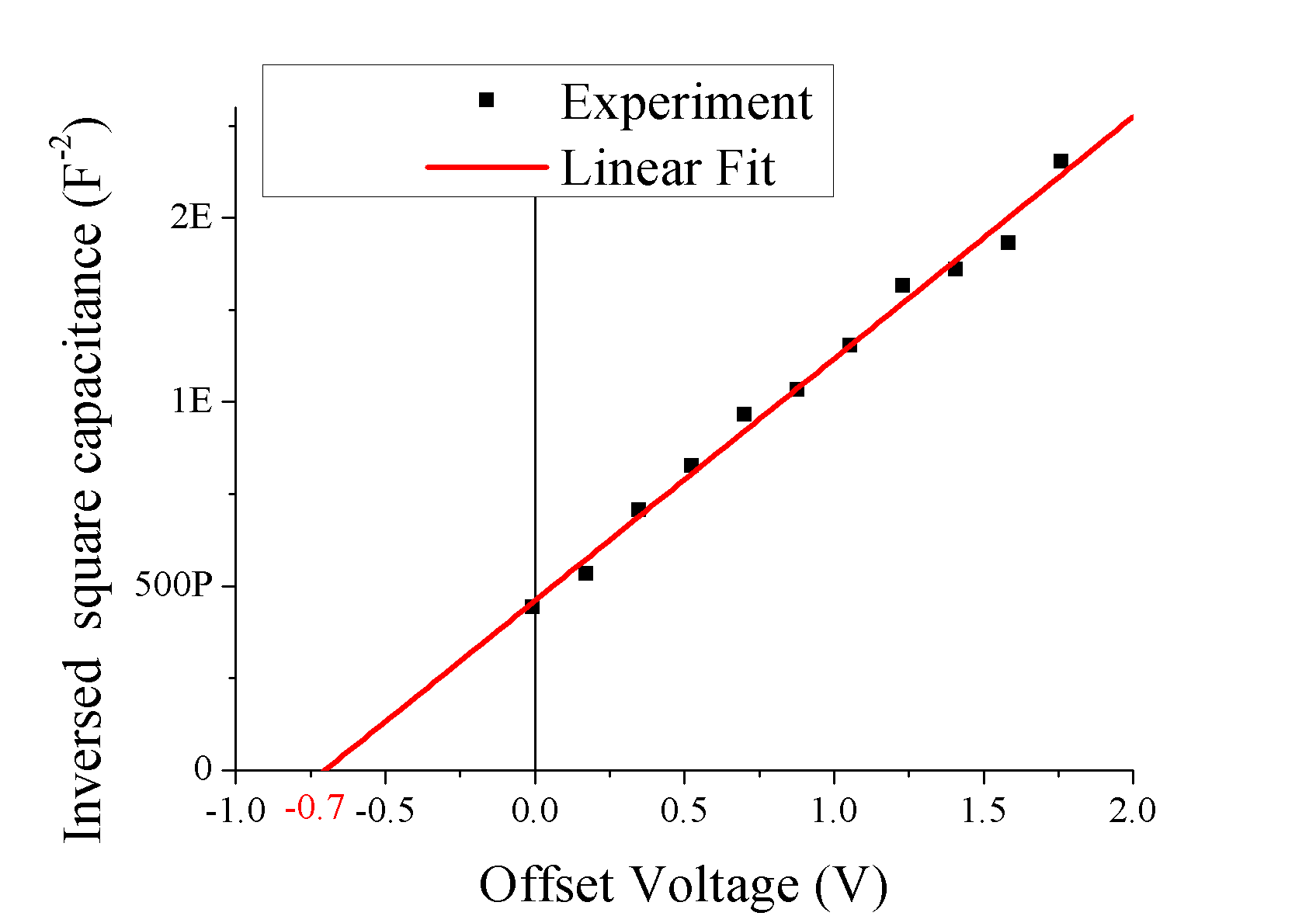}}
    \caption{Kaleidoscope of the NELM applications for various   scientific tasks. (a) and (b) Identification of simple passive RLC-circuits by NELM. (c) and (d) Measurement of junction barrier voltage using NELM for the BPW 20 RF photodiode.
     The scatter points correspond to experimental data, and the solid lines denote the CNLS approximation. For making graphs easy-to-read the experimental points were plotted using decimation with step 100 Hz. One can see that NELM can be successfully applied for element rating determination in passive circuits as well as for non-linear electronic systems study.} 
    \label{fig:RLC}
\end{figure*}

\subsection{RLC circuit and semiconductor study}
The obtained results for RLC-circuit and photodiode are presented in Fig.~\ref{fig:RLC}. From panels (a)-(b) in Fig.~\ref{fig:RLC}, it is easy to see that NELM correctly fits the RLC circuits spectra, which is also justified by the resonance peak height and position agreement with the quantities $1/R$ and $1/(2 \pi \sqrt{LC})$ calculated from the NELM fit output [for RLC experiment CPE exponent in Eq.~\eqref{eq:model} was fixed at unity value]. 

The results for a more complex system, a semiconductor photodiode, are presented in panels (c-d) of Fig.~\ref{fig:RLC}. These graphs show a set of photodiode admittance spectra measured with a reverse bias voltage applied to it in the range from 0 to -1.8 V.  Analyzing obtained data with NELM we have found that the CPE exponent value for this devices is close to unity and does not significantly depend on the offset voltage. That means that the investigated pn-junctions work as ideal capacitors, and thus the considered device functioning agrees with the classical semiconductor theory~\cite{Sze}. Secondly, we have found that the spectra of the photodiode can be described by series RC circuit,  which means that its pn-junction is near-perfect because it has no significant leakage resistance across junction, which can indicate that junction defects concentration is low.  Finally, we have observed that photodiode CV-curve, calculated using CNLS, is linear, which means that its pn-junction are abrupt. Using data presented on Fig.~\ref{fig:RLC}(c-d) we have calculate the barrier diffusion potential value 0.7 V. The same value we have found from the digitized CV-plot for BPW 20 RF manual \cite{BPW20RF_Manual, WebPlotDigitizer}, which confirms the NELM applicability for studying non-linear impedance phenomena.

In summary, we can conclude that NELM not only can be used to fit the impedance spectra of electronic devices, but its application also allows extracting their key electrical and structure characteristics.

\subsection{Cells research}
Contrary to the widespread CNLS software products, NELM was developed with the ability to analyze the non-stationary impedance data, which were stored as chronological set of files. This feature was introduced to NELM, especially for processing bio-impedance data, in particular for studying living cells using the electrical cell-substrate impedance sensing (ECIS) concept.  The original ECIS technique consists of the impedance measuring of the microelectrode covered by cells, and conventionally for cells state determination in ECIS used raw impedance data in the format ``frequency $v.s.$ impedance value'' \cite{cui2017real, das2012electrical} or in the Cell Index format \cite{ke2011xcelligence, golke2012xcelligence}, which is the normalized impedance at specific frequency. In this section, we will show the advantages of the combination of CNLS with ECIS by demonstration of the NELM application for ECIS data.  To do this, we have used HeLa cells \cite{masters2002hela}, which were obtained from the Institute of Cytology RAS, and for experiments they were incubated in multielectrode arrays in 5\% CO$_2$ atmosphere at 37$^\circ$C in DMEM medium (Biolot, Russia) overnight. Before experiments, the cells' medium was replaced with a phosphate saline buffer (PBS, Biolot, Russia). 
The chronological impedance spectra were collected once at 500 ms in the range 20 Hz - 40 kHz (2 Hz resolution) with setup, discussed in Refs.~\cite{Stupin2017, stupin2021bioimpedance}.

\begin{figure}
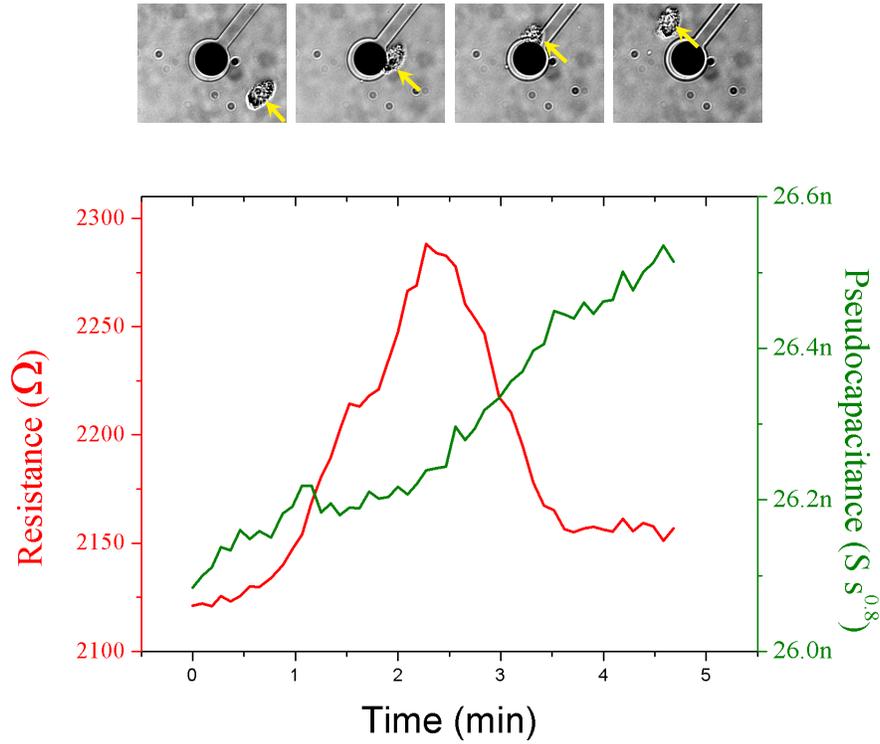

    \begin{center}
        \begin{tikzpicture}
    \node at (0,0) {\includegraphics[scale=0.25]{Trunc_4.jpg} \includegraphics[scale=0.25]{Trunc_12.jpg} \includegraphics[scale=0.25]{Trunc_16.jpg} \includegraphics[scale=0.25]{Trunc_20.jpg}};
    \node at (-0.25,-5){\includegraphics[width=12 cm]{Res_Cap_.png}};
    \def\x{-3.45} 
    \def\y{-0.5}
    \draw [-stealth, Yellow, ultra thick] (1.25+\x,-0.27+\y)--(1+\x,0+\y);
    \def\x{-1.7} 
    \def\y{-0.08}
    \draw [-stealth, Yellow, ultra thick] (1.3+\x,-0.28+\y)--(1+\x,0+\y);
    \def\x{0.25} 
    \def\y{0.3}
    \draw [-stealth, Yellow, ultra thick] (1.3+\x,-0.28+\y)--(1+\x,0+\y);  
    \def\x{2} 
    \def\y{0.5}
    \draw [-stealth, Yellow, ultra thick] (1.3+\x,-0.28+\y)--(1+\x,0+\y);    
    \end{tikzpicture}
    \end{center}
    \caption{The demonstration of the NELM applicability for the cell biophysics research. In this experiment we have measured the impedance of the bioelectrode when the living cell is floating above it (denoted by arrows). It can be seen that obtained by CNLS electrodes resistance is increasing synchronously with cell movement, which is in accordance with Gi\ae ver-Keese model (R-CPE equivalent circuit usage for cell/electrode interface). At the same time, the electrodes pseudo-capacitance is stable to the cell's movement, which is also agree with R-CPE model for bioelectrode, because the pseudo-capacitance effects should be significant when cell interact with electrode on the distance about 8 \AA~(Debye length for 0.9\% NaCl solution), which  is lower than ``electrode -- floating cell'' distance.}
    \label{fig:Follow_cell}
\end{figure}

To demonstrate the advantage of NELM for cell research, we have show the experimental proving for Gi\ae ver-Keese statement above resistive action of the cell influence of the electrode/electrolyte impedance \cite{Giaever_book, Giaever2,stupin2021bioimpedance}. To do this, we captured the evolution of the microelectrode's impedance near which the single cell detached from the MEA dish is floating.
As can be seen on the Fig.~\ref{fig:Follow_cell}, NELM-approximation had successfully detected it, reacting to this process by raise and drop of the resistance, which is caused by electrical current blocking by cells membrane, as predicted by Gi\ae ver-Keese model.  It is even possible to pinpoint the moment when the cell was fully above an electrode with the help of resistance $v.s$ time plot and data from photographs, which correspond with each other. At the same time, the pseudo-capacitance of the electrode is stable during experiment, which proves that living cell placed in proximity to electrode acts as passive resistor. Moreover, in contrast to resistive part, which reflect the biological phenomena,  obtained   by NELM data include also the pseudo-capacitive part, which reflect the electrochemical structure of the microelectrode (TiN)/electrolyte(0.9\% NaCl) interface. In the above experiment, we found that such a system acts as a CPE element \cite{About_C_Dispersion, About_C_Dispersion_2} with a non-ideality parameter value of 0.8, which indicates that diffusion phenomena and surface inhomogeneity effects significantly determine the mechanism of current flow through it \cite{stupin2022copper, panov2020situ, de1965influence, liu1985fractal,itagaki2010complex,stupin2018take}.

The examples presented in this section argue that NELM is a comprehensive tool for solving a wide variety of scientific problems, from electrical schemes testing to living matter investigation.

\FloatBarrier

\subsection{AI-based adaptive filtering approach for noise combat}
Bioimpedance measurements are often performed in noisy environments because low voltage and current values are required for the safety of biological objects. One of the software methods for noise suppression is impedance spectroscopy based on adaptive filtering (AF) (see Tab.~\ref{tab:models}), which is implemented in NELM in the procedures \verb!AF_get_w.m! and \verb!M_AF!. The first one is used for AF training (obtaining the Wiener solution using the Gaussian normal equation method \cite{Stearns}), and the second one is used to calculate the AF admittance spectrum. The main pre-tuning parameter for the AF model is the number of weighting coefficients (or the order of the adaptive filter, WC), which can be set in the \verb!Channels.Model_Options.ell_n! (voltage WC) and \verb!Channels.Model_Options.ell_d! (current WC)  fields. As input to the training process, the \verb!AF_get_w.m! function takes as arguments the measured time-domain excitation voltage and current response sequences. The third parameter to the \verb!AF_get_w.m! procedure is the \verb!Model_Options! structure, which controls the training process. $E.g.$, if the \verb!Model_Options.exact_Gram_Calc! field is \verb!true!, then AF will exactly calculate the Gram matrix for the system of normal Gaussian equations, which for high orders of AF can be a time-consuming process. In contrast, if this field is \verb!false!, then the Gram matrix will be calculated approximately, but this process will take no longer to evaluate than the FFT.
To save the AF data in ASCII format, the user can set the \verb!options.export_to_ascii=true! field, after which NELM will export the spectra during training to \verb!data*_AF.dat! files in the folder where the voltage and current time domain response sequences are stored. For further CNLS processing of the AF spectra, the user can set the \verb!I_select_file_type_number=2!  variable (in the original source-code version 1.0 beta) at the beginning of the main \verb!NELM.m! script before running NELM, which tells NELM that it should now provide CNLS processing with the AF data in ASCII format as input. In other words, to process the AF, the user must call NELM twice: once to obtain the denoised AF spectra and once to analyze these CNLS spectra. Note that the theory for AF impedance spectroscopy in Ref.~\cite{Stupin2017} is based on the assumption of the infinite length of current response and excitation voltage, so to prevent the overlearning (interpolation effect)  it is necessary to have sufficient excitation voltage and current response length for selected AF order.

\begin{figure}
    \centering
     \subfigure[]{\includegraphics[width=0.45\textwidth]{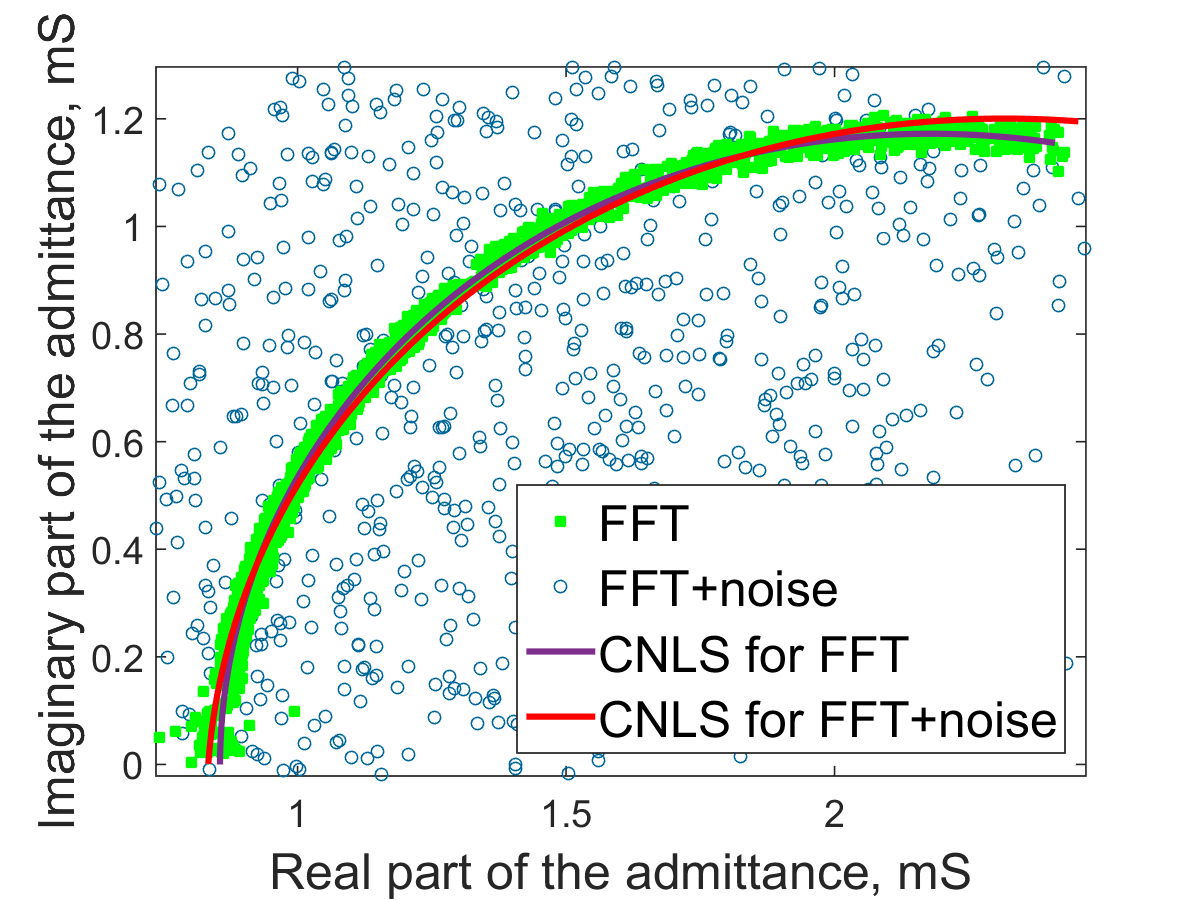}}  \subfigure[]{\includegraphics[width=0.45\textwidth]{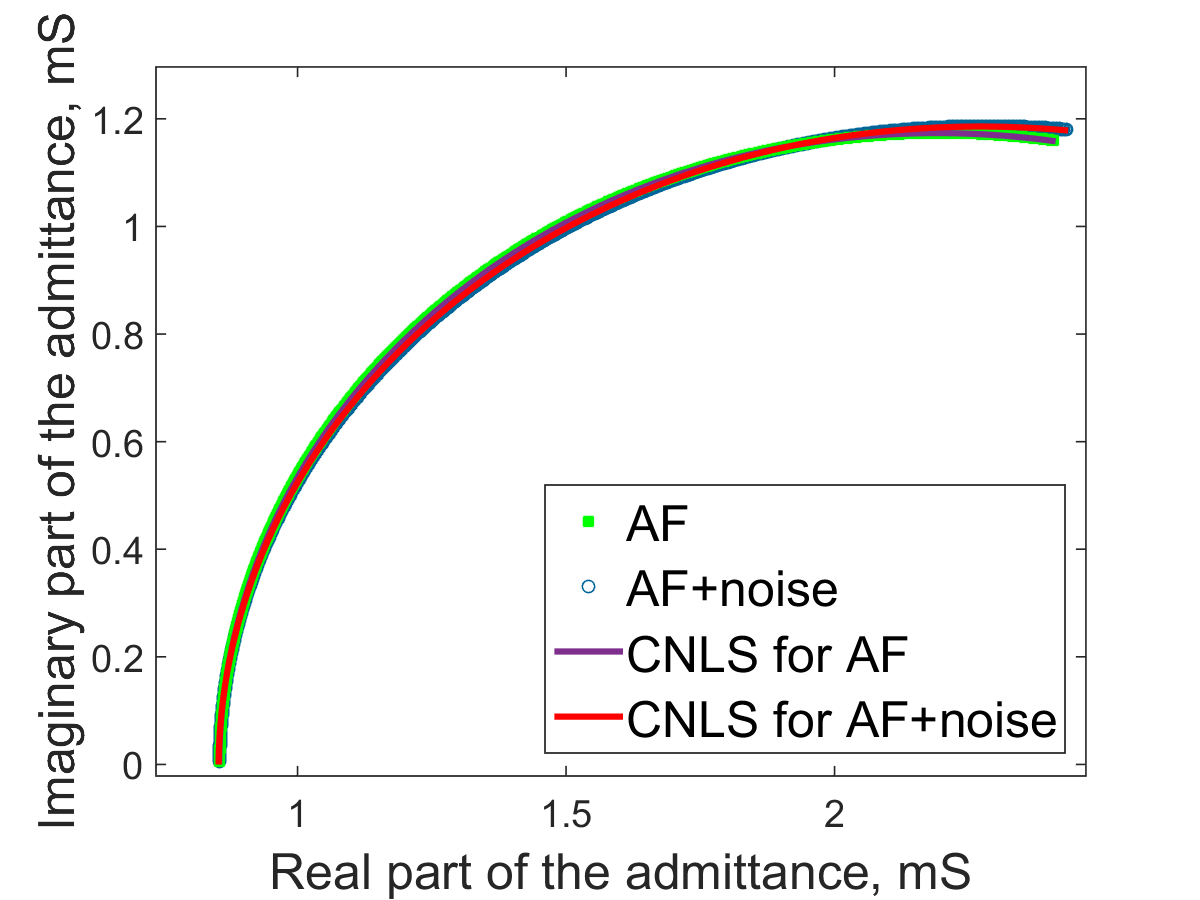}} \\
\subfigure[]{\includegraphics[scale=0.33]{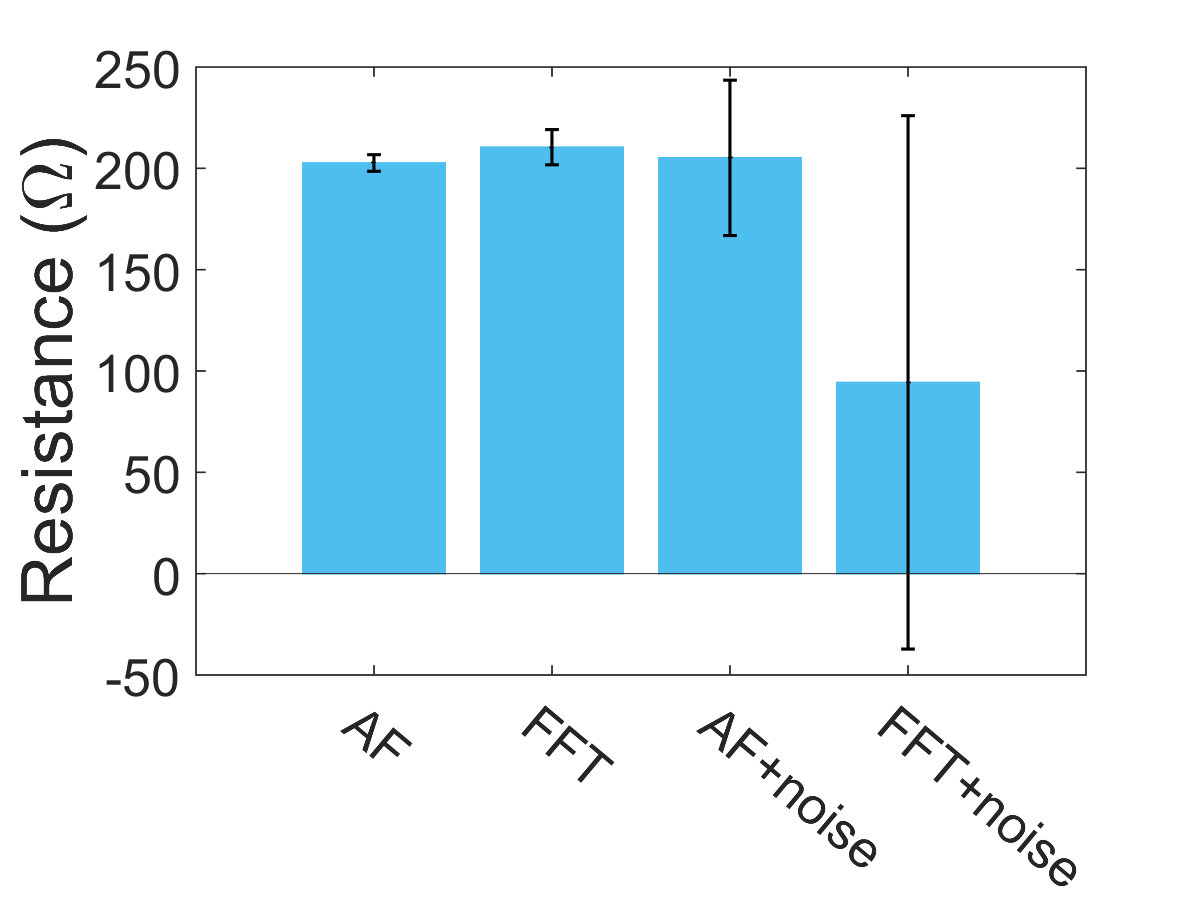}}~\subfigure[]{\includegraphics[scale=0.33]{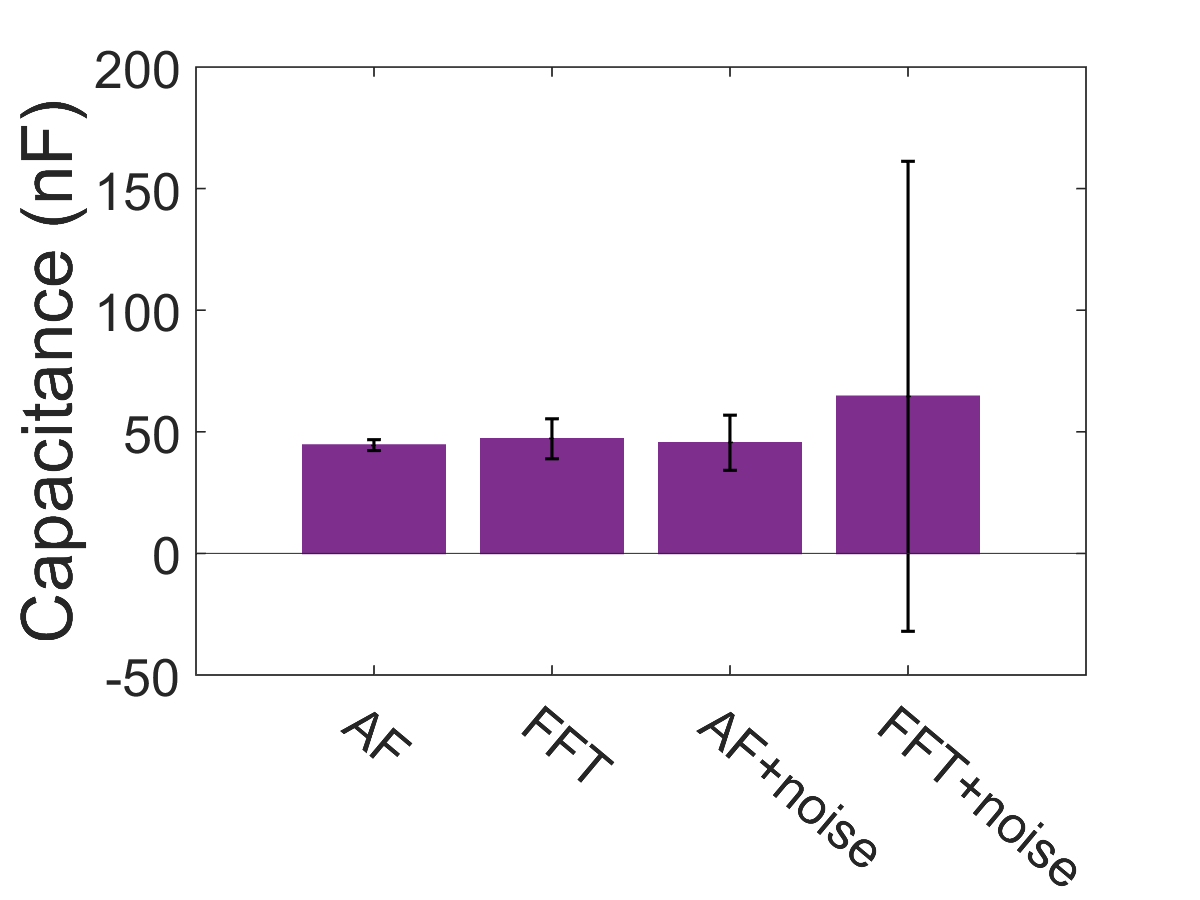}}~\subfigure[]{\includegraphics[scale=0.33]{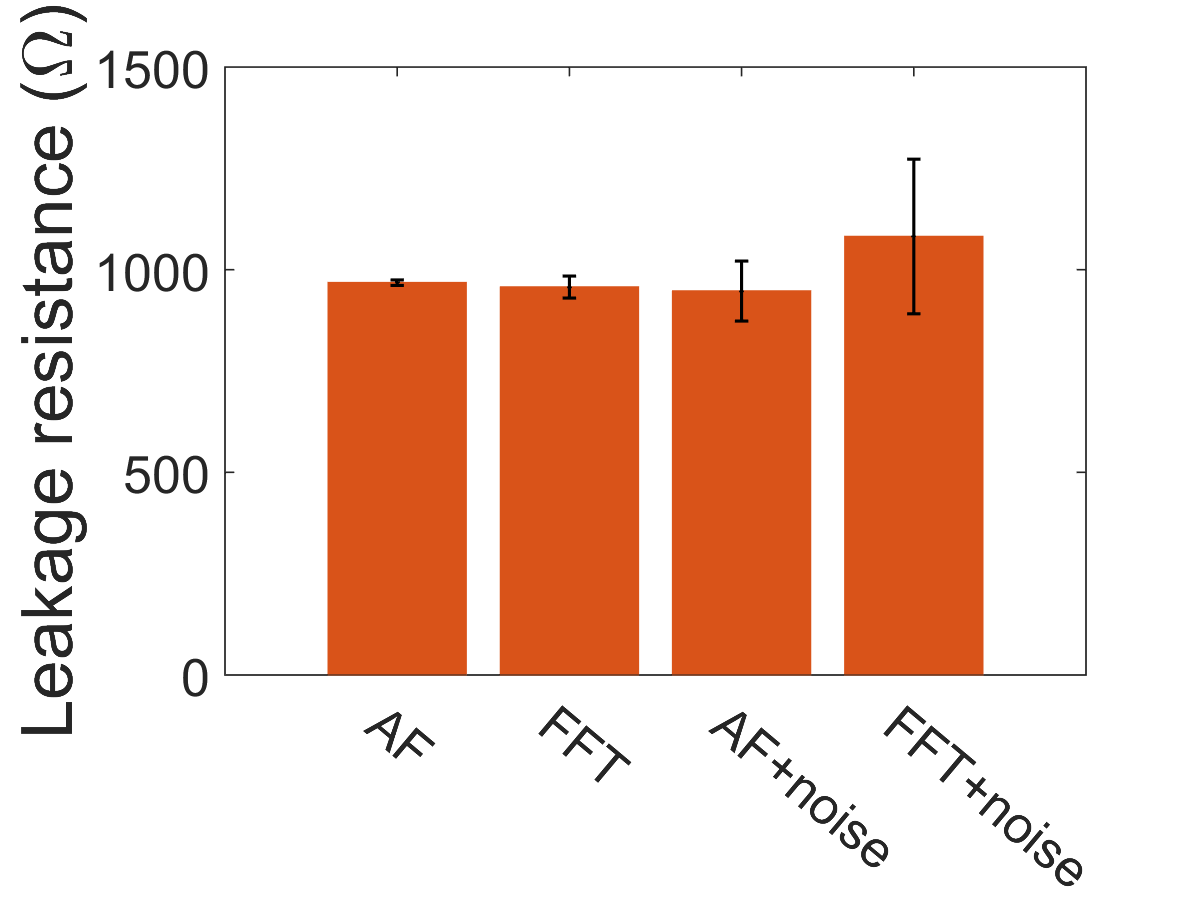}}

    \caption{Demonstration of the AF impedance spectroscopy noise immunity. Here green dots are corresponds to the experimental input data with high SNR, blue dots corresponds to the spectra obtained with low SNR, red and magenta lines are corresponds to the spectra CNLS fit with Lapicque model (Tab.~\ref{tab:models}).  Panel (a) show example of the FFT-approach usage, panel (b) corresponds to AF-technique. Lower panels  presents the obtained Lapicque scheme elements ratings. For obtaining 99.9\%-confidence intervals the statistics were collected over 10 experiments. The presented data clearly demonstrate, that AF is much stable to external random noises, which make this approach is promising for 	wearable biosensors or implants. }
    \label{fig:AF_vs_FFT}
\end{figure}

The example of the AF noise-canceling effect is presented in Fig.~\ref{fig:AF_vs_FFT}, on which are depicted immittance spectra of the frequently used in biology Lapicque circuit (see Tab.~\ref{tab:models}), with and without external white noise numerical addition.  The signal-to-noise ratio (SNR) -6 dB is used for noisy data simulation.
One can see that in the noise absence the CNLS results for both AF and FFT approach are identical, however, after noise addition only AF is able to estimate the elements rating because the FFT-estimation has error much higher than the mean rating values.
Thus, the build-in the NELM AF-based approach increases the chance of calculating the correct CNLS fit even in the case where SNR is negative, which can be observed in the real-life bioisensing application. It should be stressed that calculated model spectra are similar both for AF and FFT for both high and low SNR, while CNLS rating are significantly different for FFT-approach. This ``mirage'' effect is very dangerous for fit-quality visual estimation. The ability of the AF-technique to detect correct elements ratings confirms the necessity of using noise-suppression methods in impedance spectroscopy.
\newpage

\newgeometry{left=0.5cm,right=0.5cm} 
\restoregeometry
\FloatBarrier

\section{Conclusion and outlooks}
In this study, we have developed NELM software, which allows us to perform powerful impedance spectra analysis, including the CNLS-approach, statistical treatments, noise canceling adaptive filtering, as well as  raw spectra visualization.

We tried to make our software as flexible as possible, allowing the operator to be in full control of the approximation process. 
To be specific, the user can not only choose  build-in  approximation protocols, but also they can construct their own model schemes, input/output data files, and approximation types by included in NELM package set of the constructing tools. 
Also, we tried to sculpt NELM as user-friendly software, and for this reason in addition we tried to equip its messages with human-like quirky language with some sense of humor. Possibly, in the future we will equip it with more progressive speaking features based on AI.

Implemented in the modern \textsc{MatLab} language with the usage of the open source ideology, the NELM is the next-step-generation software for impedance data analysis, which combines the modern parallel computing opportunities with well-tuned \textsc{MatLab} algorithms and abilities of symbolic calculus.
In addition, even though NELM was originally created in order to analyze bioimpedance, we have proved that it is capable to solve same tasks for the other fields of science, from semiconductor studies to electrochemistry. Thus, the results of our study can find an applications practical biology, experimental medicine, semiconductor technology, electronic industry, electrochemical and biosensing devices. We believe that NELM will be helpful for fundamental biology and bioelectronic research, as well as for solutions of relevant problems of healthcare and environmental studies. We will try to make the NELM updates according to the last scientific discoveries and challenges in the art of approximation.

\section*{Acknowlegement}
The authors are grateful to Kraevskya A.A. for valuable programming advise, Yakovleva L.E., Dubina M.V. for comprehensive assistance and support. This study was carried out with the support of the Ministry of Science and Higher Education (Project № FSRM-2024-0001).


\end{document}